\documentclass{aa}  

\usepackage{graphicx}
\usepackage{txfonts}
\usepackage{booktabs}
\begin{document}

   \title{The lithium-rich giant stars puzzle: New observational
trends for a general-mass-loss scenario}

   \subtitle{}

   \author{R. de la Reza
          \inst{1}
                   }

   \institute{Observatorio Nacional (ON), MCT, Rua José Cristino 77, Sao Cristovao, Rio de Janeiro -Brazil\\
              \email{ramirodelareza@yahoo.com}     
             }

   \date{}
% \abstract{}{}{}{}{} 
% 5 {} token are mandatory
 
  \abstract
    {The existence of one percent of lithium-rich giant stars
among normal, lithium-poor giant stars continues to be poorly
explained. By merging two catalogues - one containing 10,535 lithium-rich
giant stars with lithium abundances ranging from 1.5 to 4.9\,dex, and the
other detecting infrared sources - we have found 421 clump giant stars 
and 196 first-ascending giant stars  with infrared excesses indicating
stellar mass losses. The clump stars are the most lithium-rich.
Approximately 5.8 percent of these stars appear to episodically lose
mass in periods of approximately $10^{4}$\,years or less, while the remaining
stars ceased their mass loss and maintained their lithium for
nearly $10^{7}$\,years. We propose a scenario in which all giant stars with  
masses below two solar masses undergo prompt lithium enrichment
with mass-ejection episodes. We suggest that the mass loss results
from internal angular-momentum transport. It is possible that a
transitory instability, perhaps of magnetic origin, rapidly transports the
nuclear material responsible for the lithium enrichment to the stellar
surface and triggers shell ejections. Additionally, the strong mass loss
in some lithium-rich stars during their evolution activates their
chromospheres, as observed in ultraviolet spectra. Furthermore,
intense episodical mass losses in these stages led to the observable
formation of complex organic and inorganic particles, as detected in
near-infrared spectra. In contrast to first-ascending giant stars, helium
flashes during the clump can contribute to additional lithium
enrichment alongside the aforementioned process. The combination of
these two lithium sources may explain the much higher observed
lithium abundances in clump stars, as well as their observed infrared
excesses. If our scenario --based on a universal and rapid lithium
enrichment episode process-- is correct, it could explain the rarity of
lithium-rich giant stars.     
  }


   \maketitle
%
%-------------------------------------------------------------------

\section{Introduction}
Since the first Li-rich K giant star was serendipitously discovered by
\cite{Wallerstein1982}, with a lithium abundance of 2.95\,dex, a
complete explanation of this phenomenon has not appeared up to now.
This situation is commonly referred to in the literature as part of the
'lithium giant stars puzzle'. In fact, if the standard model of stellar
evolution can predict lithium abundance values below 1.5\,dex, larger
values than this require non-standard physical scenarios. After 1982,
several Li-rich stars were discovered, primarily by larger surveys such
as the LAMOST survey \citep[][hereafter GA19]{Gao2019}, the GALAH survey 
\citep[][hereafter MA21]{Martell2021}, the GAIA-ESO survey \citep{Smiljanic2018}, and \cite{Casey2016}, among others. Even though several thousands of Li-rich red
giant stars have been detected today, they still represent only 1--1.5\% of
the Li-poor red giant low-mass stars with masses between $\sim$\,0.8 and $\sim$\,2\,M$_\odot$. The absence of an explanation for this permanent fraction of $\sim$\,1\%
can be considered as the other part of the puzzle.
Regarding the proposed scenarios to try to solve this Li giant stars
problem, a dedicated paper on the collection of them has been
presented by \cite{Yan2022}. As a general view, these scenarios are
divided into a stellar internal origin of the Li enrichment and an external
one, or a mixture of them. The internal ones use the known Cameron-Fowler mechanism \citep{Cameron1971}, based mainly on the
process in which the produced $^{7}$Li is transported to the surface
convective stellar envelope, where it can be observed. The external
mechanisms require an external source of Li as planets or sub-stellar
objects, which are ingested by the external stellar red-giant convective
zone \citep{Siess1999,Carlberg2012,AguileraGomez2016,Stephan2020,SoarezFurtado2021,AguileraGomez2022,AguileraGomez2023,Tayar2023,Behmard2023}.
However, these external mechanisms require, as a validity test, not only that the planetary or sub-stellar source of Li appears as an excess in
the star, but also the presence of other elements such as Be and B. Nevertheless,
\cite{Castilho1999} and \cite{Melo2005} did not detect any
observed excess of Be in Li-rich giant stars. The same is also the case
for the boron element, which is the hardest to destroy in internal
stellar layers, making it the most severe test for the engulfing scenario.

The boron feature in the UV \textit{Hubble} spectra at 1200 \text{\AA} does not show the
presence of any excess of boron in some very Li-rich and Li-rich giant
stars \citep{Drake2018}. It must also be noted that the engulfing
mechanism cannot explain the existence of what are called the very
Li-rich giant stars with Li abundances larger than 2.2 dex \citep{AguileraGomez2016}. Concerning the mixture cases, \cite{Casey2019}
proposed that a tidal spin-up effect produced by a binary companion
\citep[see also][]{Costa2002} can induce sufficiently fast mixing in the
giant stellar interior, which is capable of driving $^{7}$Li production. This prediction of
the presence of binary stars is especially invoked for the red-clump (RC) giant stage of
evolution. However, a recent work by \cite{CastroTapia2024},
investigating the radial velocities of quite a large sample of 1400 giants,
does not find evidence of a binary nature of Li-rich giant stars, at least
for RC giant stars \citep[see also][]{Tayar2023}. Apart from these examples, there are other more specific cases of Li production, such as a
merging between a red giant star and a He white dwarf \citep{Holanda2020,Zhang2020}.
The study of Li-rich and very Li-rich giant stars, which means up to Li
abundances of $\sim$\,5\,dex, requires an approach that goes further than the
standard and non-standard models known in the literature. Concerning
a standard study in which the stellar convective motions are the only
mixture mechanism, a recent, different approach was taken by \cite{Li2024}
showing that in the absence of a Li-depleting mechanism, and
progenitor stars with Li abundances that exceed 3.3\,dex, the majority of
stars will be Li-rich. However, in this approach the Li problem brings a
new problem, which is the need to search for an unknown and very
efficient extra Li-depletion process that affects the largest and major part of
the giant stars, which are in fact Li-poor. Concerning non-standard
models, such as those based on thermohaline instability and rotation-induced mixing processes \citep{Charbonnel2010}, these are
not able, at least for first-ascending giant (i.e. RGB) stars, to produce larger increases in Li
abundances. As a result of this work, where we investigated the internal Li-enrichment process in low-mass giant stars, we propose that the entire
Li-enrichment scenario is a consequence of the evolution of a single
star. This approach has no limitations regarding the observed Li
abundances between 1.5 and $\sim$\,5.0\,dex.

This paper is structured in the following way. Section 2 is devoted to all
aspects referring to the IR (infrared) excesses found here for Li-rich giant stars. In
Sect. 3, a general scenario with the problems involving the angular
momentum and the corresponding nuclear reactions are presented.
Section 4 is dedicated to the general Li properties in giant stars. Finally,
a discussion and conclusion are presented in Sect. 5.
  
\section{Measurements of the infrared excesses}

To our knowledge, few works have been dedicated to the measurements
of IR excesses for a larger number of giant stars such as MA21,
\cite{Mallick2022}, and \cite{Sneden2022}, among others. For
our research here, we are guided by the work of MA21
 because it provides more details of their search for IR
excesses. MA21 considered, as they call it, 'clean' WISE detections,
which are sources that do not show image confusion. This is done
following the instruction in the WISE manual, with the WISE cc\_flags
confusion flag set to 0000, corresponding to the four W1 (3.4\,$\mu$m), W2
(4.6\,$\mu$m), W3 (12\,$\mu$m), and W4 (22\,$\mu$m) magnitude colours. Additionally, a second condition was used in which the ph\_qual photometric quality flag is set to A for W1 and W4.

\begin{figure}
    \centering
    \includegraphics[width=.99 \hsize]{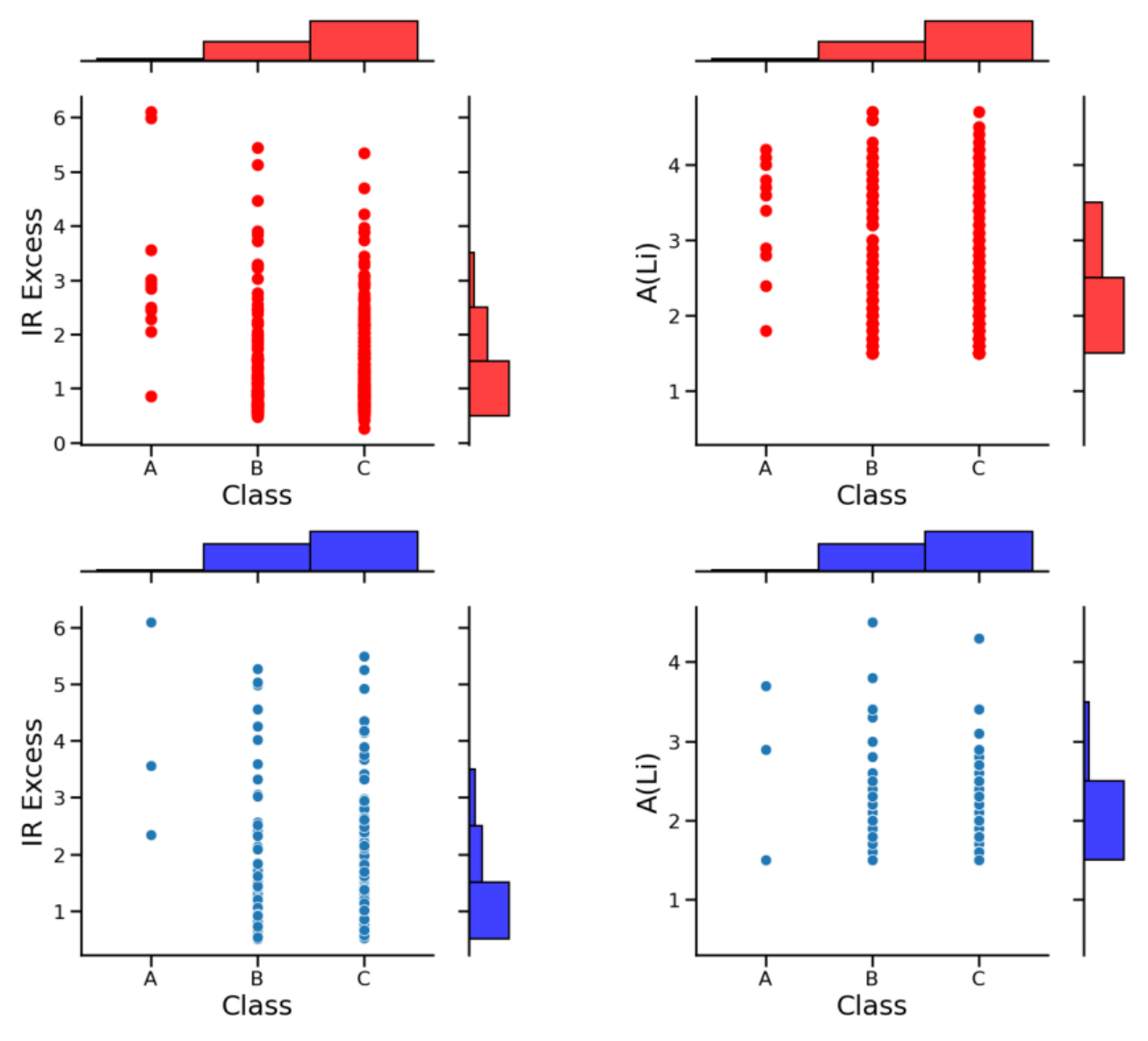}
    \caption{
    Distribution of RC and RGB giant stars in different classes of signal-to-noise ratios A, B, and C for the IR excess measured by the difference of WISE bands W1 - W4 in magnitudes related flux densities expressed in Janskys (Jy) (first column) and for the Li abundance in dex  (second column). Histograms of these distributions are presented in the right vertical and horizontal scales.
    }
     \label{fig1}
\end{figure}

From a set of 1862 giant stars, mainly from the GALAH catalogue, with clean WISE detections, they detected only three (two RC stars and one RGB star) with larger excesses W1--W4 larger than 0.5 (see their Figure 9). We believe, as we show later, that
this very low number is the result of the extremely strict detection
conditions used by these authors, considering only the mentioned A
quality flag. Therefore, we decided to explore the whole LAMOST
catalogue containing 10,525 Li-rich giant stars (GA19) under less strict WISE detection conditions. For this purpose, we retained the first confusion flag, keeping sources with the 0000
confusion condition, but gradually relaxed the ph\_flag, exploring
conditions other than A (which represents the signal-to-noise ratio S/N
$>$\,10 ) to levels B (S/N\,$<$\,10) and finally to level C (S/N\,$<$\,3).
We eliminated, a
priori, all IR sources presenting upper-limit values in W1 and W4. The
main condition is that all of our accepted W1 and W4 values fulfil the
measure of W1--W4 of being larger than or equal to 0.5 in order to be
considered as good indicators of the infrared excesses. The detailed
conditions of acceptance or refusal of sources are given in the
instruction manual of WISE. After, we will compare the representativeness
of the results obtained with the three classes A, B, and C.
As a final result, we found 421 RC stars and 196 RGB stars, with a total
of 617 stars presenting IR excesses and fulfilling the WISE magnitude
colour difference of W1--W4 $\geq$\,0.5 considered here as a measure of the
presence of an IR excess. The numbers' distribution with the three
different classes is 14 stars in class A, 208 in class B, and 395 in class
C. A comparison with the work of MA21 for class A using only
class A in a sample of 1862 stars gives the following rates: for MA21, it
is 3/1862 or 0.16\%, whereas in our case, we have 14/10535 or 0.13\%, which is a
quite similar result. For case B, we obtain 1.97\%, and for case C, 3.7\%.
Due to the good distribution representativeness of values of A(Li) and
IR excess for classes A, B, and C, as displayed in Fig. \ref{fig1}, we
considered in this work that all measurements obtained with classes A,
B, and C are reliable to proceed for a general analysis using these
parameters. All results of the measurements of the infrared excesses
contained in the LAMOST catalogue of Li-rich giant stars are presented in
Table~\ref{tab:table1} for 421 RC giant stars and for 196 RGB giant stars.
Table~\ref{tab:table1}
contains all the stellar parameters as presented in the LAMOST catalogue
of Li-rich giant stars, to which we added four columns
corresponding to the IR excess measurements.

The parameters presented in Table~\ref{tab:table1} are the name of the object; coordinates;
$T_{\rm eff}$; log g; [Fe/H]; A(Li); IR excess; CC\_flag; Ph\_flag; class; and the
classifications A, B, and C with the respective  numbers 1, 2, and 3.
We note that our mentioned working catalogue, LAMOST, contains, Li-rich
giant stars for the following stellar parameters: $T_{\rm eff}$ between
approximately 4000\,K and 5500\,K, log g between approximately 1.0 and
3.5, and [Fe/H] between approximately -1.7 and 0.4.

\begin{table*}
\centering
\caption{Stellar parameters of RC and RGB giant stars presenting IR excesses. }
\label{tab:table1}
\begin{tabular}{ccccccccccc}
\toprule
OBJECT & Coordinates &$T_{\rm eff}$ (K) &log g&[Fe/H] &A(Li)& IR& CCflag& Phq& Note & Stage\\
\midrule
J000108.96+072932.9 & 00:01:08.96   +07:29:32.9 &  4731 &  2.5  & -0.2 &  3.6 &  1.067  & 0  & AAAC &  3 & RC \\
J000151.65+265848.4 & 00:01:51.65   +26:58:48.4 &  5071 &  2.5  & -0.5 &  4.7 &  1.208  & 0  & AAAC &  3 & RC \\
J000851.01+523009.3 & 00:08:51.01   +52:30:09.3 &  4797 &  2.5  & -0.1 &  2.3 &  1.553  & 0  & AAAC &  3 & RC \\
J000920.04+563644.1 & 00:09:20.04   +56:36:44.1 &  4830 &  2.4  & -0.3 &  3.9 &  1.654  & 0  & AAAC &  3 & RC\\
J001154.03+564857.0 & 00:11:54.03   +56:48:57.0 &  4785 &  2.4  & -0.2 &  1.8 &  0.584  & 0  & AAAC &  3 & RC \\
... & ... &  ... &  ...  & ... &  ... &  ...  & ...  & ... &  ...\\
\bottomrule
\end{tabular}
\tablefoot{The table includes, in order, the stellar identification, right ascension (RA), declination (DEC), effective temperature ($T_{\rm eff}$), surface gravity (log g), metallicity ([Fe/H]), lithium abundance (A(Li)), IR excess (IR), the WISE confusion flag, and the WISE photometric flag. The notes indicate the classification of IR sources based on the photometric flag (see text) and  evolutionary stage (RC or RGB). This table is shown in part; the complete version is available online as a Vizier catalogue.
 }
\end{table*}

\subsection{The use of log g to separate RGB and RC giant stars}

Distinguishing between RGB and RC giant stars is not an easy task.
The best way to separate them requires the use of astereoseismology.
This is due to the fact that RGB and RC stars have different internal
structures, and they exhibit different signals of seismical oscillations  \citep{Bedding2011}.
Normally, without this tool, this distinction is generally made using a
standard methodology to obtain the stellar gravities (log g). The main
known problem is that if RC giant stars are concentrated around certain
specific values of log g, this is not the case for RGB giant stars' log-g
values, which occupy a larger range of log-g values, even comprising
those corresponding to RC values. Under these conditions, disentangling
RC and RGB classes of stars could involve some dubious cases.
We proceed to classify the most appropriate general values of
log g, guided by results based on those obtained using
asteroseismology  \citep[][hereafter YA21]{Yan2021}. YA21 provided a
uniquely large sample with a clean determination of evolutionary stages
based on asteroseismology. Under these conditions, we adopted the following
criteria: for RGB stars, log g between 2.8 and 3.5; for RC giants, log
g between 2.2 and 2.7. Adopting these conditions, we discarded
88 Li-bright giant stars that have log-g values under 2.2, for which
there are no measurements from astereoseismology (following YA21).

\begin{figure}
    \centering
    \includegraphics[width=.99 \hsize]{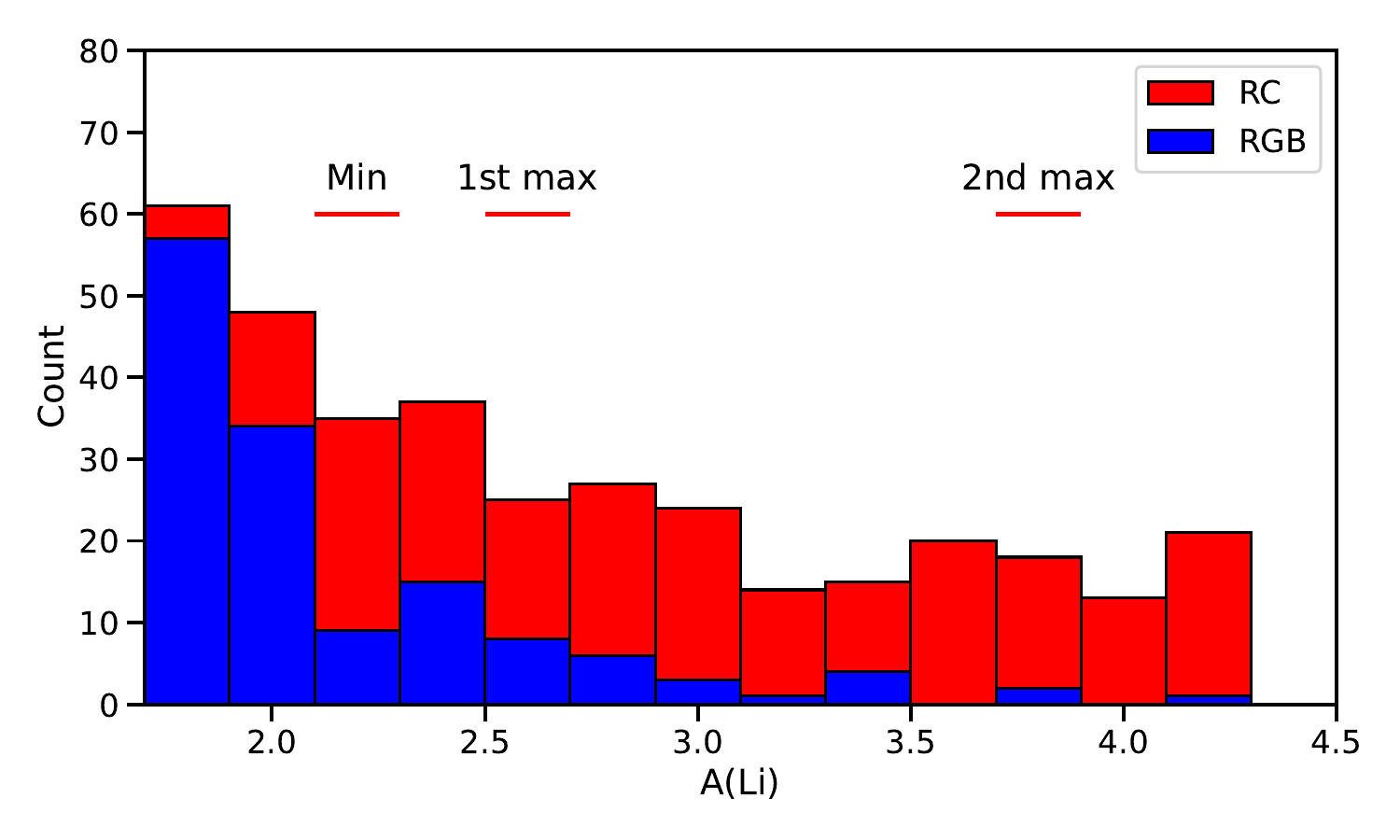}
    \caption{Counts of giant RC and RGB stars with IR excesses as function of Li abundance in dex as obtained in this work.  This figure must be
compared with a similar figure of YA21 based on asteroseismology data
(Figure 1 in YA21). Using short horizontal lines, we present 
the regions coincident with the presence of minimums and maximums
of the figure of YA21. In our work, we eliminated all bright RC
giants stars with log-g values under 2.2 for which no
astroseismology calibration exists following YA21.
    }
     \label{fig2}
\end{figure}

The counts of all our RC and RGB giants, all of them presenting IR
excesses, as functions of the Li abundance are shown in Fig. \ref{fig2}. YA21
presented a similar distribution, except for Li-rich giant stars, which do not
show IR excesses in principle and are based on asteroseismology. To compare both
distributions, in Fig. \ref{fig2} we use short horizontal lines to show the
minima and maxima of Figure 1 in YA21. This simple comparison
shows that the distributions are similar. We can then conclude that our
distribution, based on a log-g discrimination of RGB and RC stars, is
similar to that based on asteroseismology. This result indicates that
giant stars presenting IR excesses are similar to giant stars not
showing IR excesses. The same argument that both classes of objects
are similar is presented in the next Sect. 2.2, this time based on
metallicities.

\begin{figure}
    \centering
    \includegraphics[width=.99 \hsize]{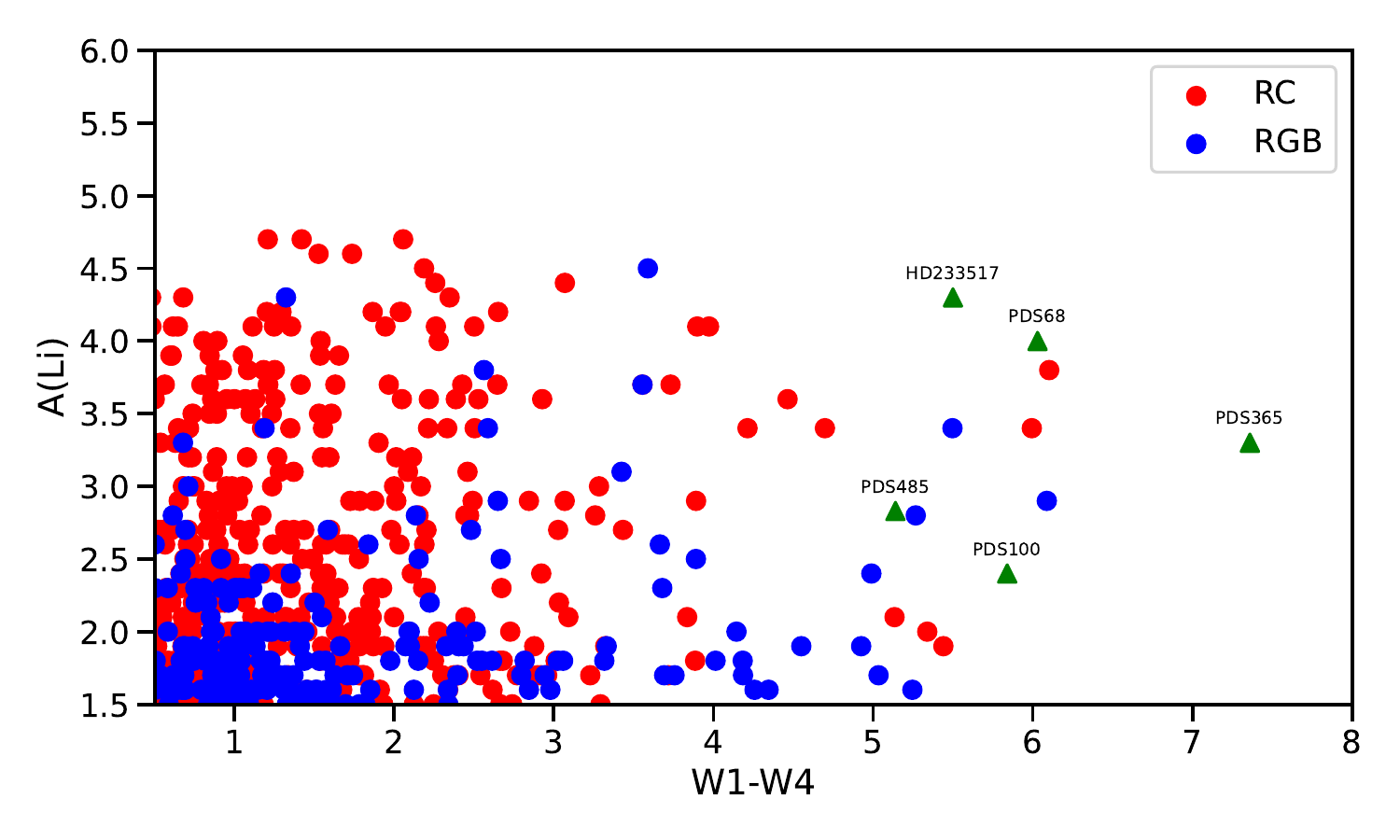}
    \caption{A(Li) in dex of 421 RC and 196 RGB giant stars
obtained in this work in function of IR excesses measured by the difference
magnitude WISE colours W1 -- W4 larger than or equal to 0.5. RC stars are denoted with red, whereas RGB stars are denoted in blue. Green triangles represent some giant stars not belonging to the
LAMOST catalogue (see text).
    }
     \label{fig3}
\end{figure}

The results of our measurements of the infrared excesses based on all
the 10535 Li-rich giants in the LAMOST catalogue are shown in Fig. \ref{fig3}.
The 421 RC giant stars whose IR excesses have been detected are
represented by red points, while the RGB giant stars are represented
by blue points. The y-axis contains the Li abundances as measured
in the LAMOST catalogue (GA19) from 1.5 up to 4.8. The x-axis contains
the measured IR excess values from W1--W4 larger than or equal to 0.5 up to
larger magnitudes. In the same figure, we superposed five Li-rich and
super-Li-rich RG and RC giant stars (green triangles) known in the
literature that do not belong to the mentioned catalogue. These
stars have the property of producing complex organic and inorganic
material in their very strong winds (see Sect. 4.3). These stars are
HD\,233517, PDS\,68, PDS\,100,  PDS\,365, and PDS\,485.
A simple inspection of Fig. \ref{fig3} shows that RC giants are dominant
among the very Li-rich giant stars. This is not the case for RGB stars,
which, apart from a few exceptions of very Li-rich objects, appear
concentrated with Li abundance values below $\sim$\,2.6\,dex. We must
note that several RGB stars with high Li abundances are suspicious due
to the mentioned difficulty of separating RGB stars from RC stars. In
these conditions, without a seismology criterion, it will be no surprise if
some RGB-classified stars with Li abundances over 2.6\,dex are
misclassified and are, in reality, RC stars.

\subsection{Metallicity of li-rich giant stars presenting mass
loss}

As mentioned before, we found a subgroup of stars containing nearly
6\% of the 10,535 studied Li-rich giant stars belonging to the LAMOST
catalogue (GA19), presenting evidence of mass loss. One
of the main arguments of this work is to affirm that this subgroup is a
transitory one, in the sense that the stellar properties of this sub-group
are the same as the rest of Li-rich stars, with the exception that they are
transitorily ejecting mass. To try to prove this fact, we now use their
invariable stellar metallic properties and compare them to the
metallicities of Li-rich giant stars not presenting IR excesses, which
means not showing indications of the existence of mass loss.

\begin{figure}
    \centering
    \includegraphics[width=.70 \hsize]{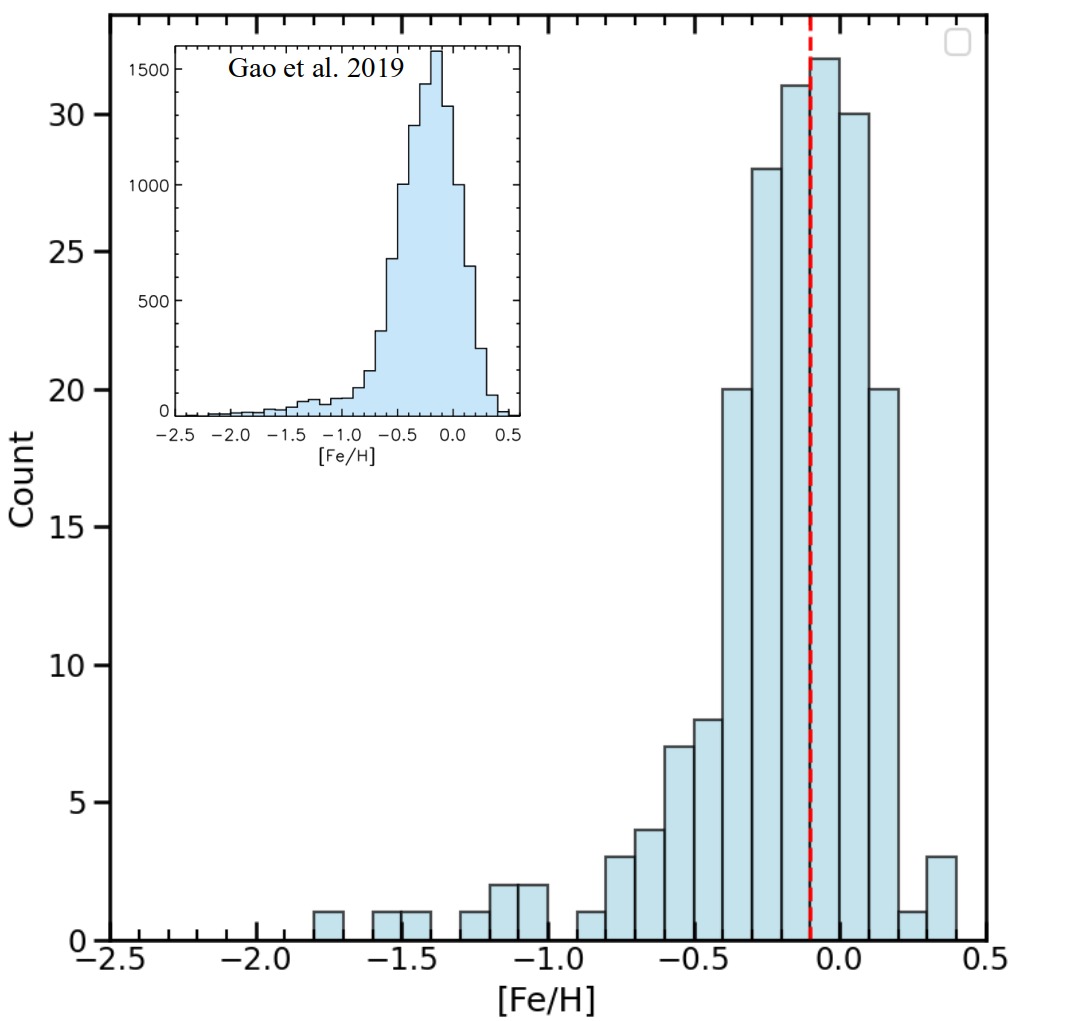}
     \includegraphics[width=.70 \hsize]{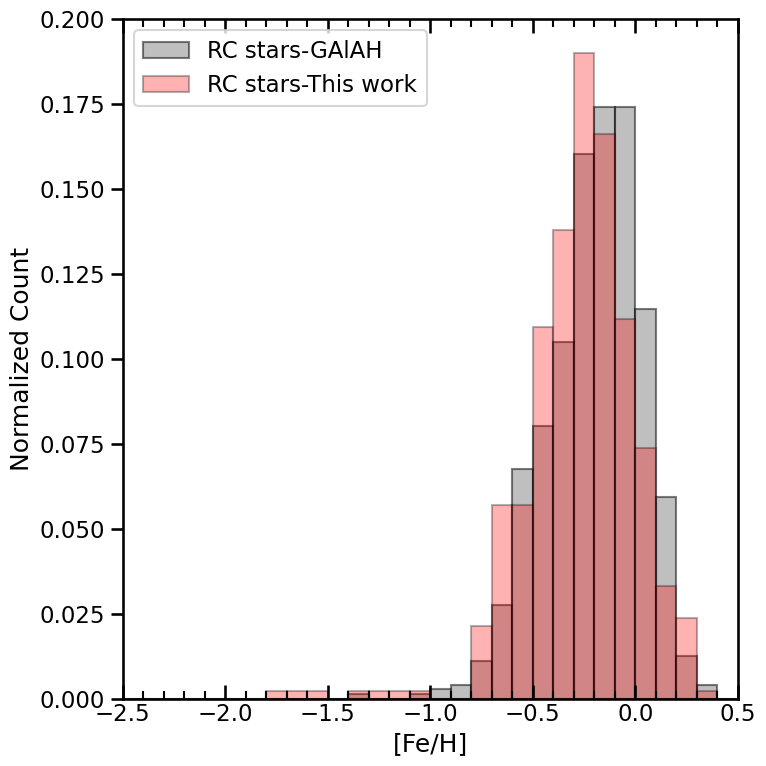}
      \includegraphics[width=.70 \hsize]{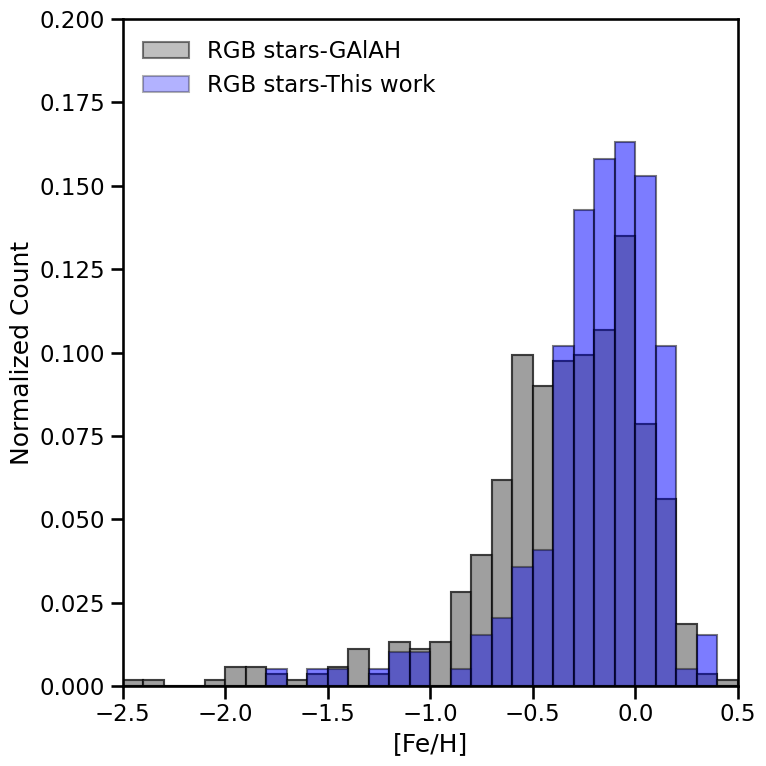}
    \caption{Comparisons of our distribution of Li-rich giant stars presenting IR excesses as a function of metallicity to the distributions of the same Li-rich giants from the LAMOST and GALAH data sets that do not show these IR excesses. }
    
     \label{fig4}
\end{figure}

In Fig. \ref{fig4}, we display metallicity histogram distributions of Li-rich
giant stars from the LAMOST and GALAH data sets and compare them
to our LAMOST Li-rich giants, this time presenting IR excesses. First,
we compare those resulting from the same LAMOST data set. In the
upper panel of Fig. \ref{fig4}, we show the GA19 distribution without
separating RC and RGB giants alongside our distribution containing IR
excesses. This results in a similar distribution, which naturally arises
because they belong to the same set of data. In the two lower panels,
the comparisons are made with the GALAH data set, this time
separating RC and RGB giants. The second lower panel, corresponding
to RC giants and in which the distributions are normalised, indicates a
rather similar behaviour, with both distributions presenting the same Full Width at Half Maximum
(FWHM). The third panel corresponds to RGB giants. Here, some
differences appear, with the FWHM of GALAH being somewhat larger
than our distribution. This difference results from a relatively flat
distribution found in GALAH from -2 $<$ [Fe/H] $<$ 0. This peculiar flat
distribution is interpreted by MA21 as eventually being due to multiple
signs of Li-enrichment processes at the core of their distribution (see
MA21). Regardless of these details concerning the attributed Li
enrichments applied only to RGB giants, their distributions of RC giants
appear similar to ours.

We conclude that
our subgroup of mass-losing giant stars, which includes both RC and
RGB stars, has a similar metallic distribution to that of all Li-rich giant
stars. In this way, we show that our subgroup is of the same nature as
all Li-rich giant stars, with the exception that they are transiently
ejecting masses.

\section{The general scenario}

With this work we present a qualitative general scenario that is valid for a
single low-mass giant star to explain the origin of the Li
enrichment observed in these stars. The main scenario follows a
universal approach in which all low-mass giant stars with masses
between $\sim$\,0.8\,M$_\odot$ and $\sim$\,2\,M$_\odot$ undergo a short episode of Li enrichment accompanied by the ejection of a circumstellar shell \citep{delaReza1996,delaReza1997,delaReza2012}. 
It is considered that the origin
of this mass loss resides in the internal angular momentum (AM)
evolution of the star. This evolution is provoked by the
asteroseismology-observed slow rotation of the stellar core (see a
review in  \cite{Aerts2019}). How this AM transfer distribution is actually
occurred in these stars is not well known. We propose that this
distribution consists of a mass transfer between the radiative upper H-burning zone and the external convective envelope \citep[see also][hereafter DLR15]{delaReza2015}.

One essential requirement in this scenario is a rapid and sporadic internal transport
of the internal mass containing the $^{7}$Li as the dominant element; this is from
the internal upper H-burning zone before it is destroyed by nuclear
proton reactions and up to the external stellar convective layer. This rapid
internal upward process is not known, and we speculate that it could be
provoked by another sporadic instability, probably of a magnetic nature.
Following the nuclear simulations presented by \cite{Yan2018}, (hereafter YA18), the
required $^{7}$Li transport time is only 47 yr for a very Li-rich giant with a Li
abundance of 4.5\,dex (see Sect. 3.1). Another important part of our
scenario consists of this mass transport being sufficiently energetic in
order to form low-mass circumstellar shells, which are observed in
these Li-rich and very Li-rich giant stars. The evolution of these rapid
shells, formation, and ejections (see Sect. 4) represents stellar winds
with much larger, episodic stellar mass losses, even with a maximum increase of
five orders of magnitude compared to the permanent stellar mass losses of
the normal Li-poor giant stars.

The proposed qualitative scenario requires that all stars are suddenly Li
enriched, at least one time, during the RGB or RC stages. Are
more of these episodes possible? This depends on the repeatability of
speculated internal instability and on the reservoir of He$^{3}$ available.
When this fuel, which originated when the star was on the main
sequence, is finished, the Li enrichment stops.

To our knowledge, only two works in the past have used the stellar AM
to study the problem of high Li abundance in giant stars: \cite{Fekel1993} and DLR15. However, in the first reference, they considered that a rapidly rotating stellar core was the
main trigger mechanism provoking the high Li abundances and
producing the IR excesses \citep[see also][]{Fekel1996}. Today, it is known
that the stellar core is, on the contrary, slowly rotating. In DLR15, this
new condition of the rotating core is taken into account, suggesting
internal mass transport as the main avenue to enrich the giant star with
fresh Li and provoke the formation of the IR excesses. This represents
the general scenario of the present work.

In fact, the detection of slower rotation in the cores of low-mass stellar giant
stars using asteroseismology techniques, compared to predictions from
standard classical models, has sparked renewed interest in the study of stellar interiors. This core spin-down suggests that there is an
efficient redistribution of internal angular momentum  transfer at
play. However, the specific mechanism for this transfer between the
radiative upper H-burning zone and the external convective zone in
actual stars has not yet been resolved. Many researchers are currently
dedicated to solving this problem, as evidenced by the references
provided in recent works \citep[see, e.g.][]{Meduri2024,Moyano2023,Dumont2023,Denissenkov2024}. Both magnetic
rotation instability and Tayler instability, driven by differential rotation in
giant stars, are competing processes. \cite{Meduri2024} also
introduce the important transfer of chemicals from the internal regions
to the star's surface, creating another avenue for future developments
in this field.

In this context, one partial requirement to try to solve the Li puzzle
consists of examining the rapid upward motion of the internal material,
where $^{7}$Li is the dominant element. This motion considers the travel
from the upper H-burning region up to the external convective envelope.
\cite{Yan2018}, using the conveyor-belt mechanism
\citep{Sackmann1999}, made all the necessary nuclear-reaction
simulations in order to result in a very Li-rich giant star with a Li
abundance of 4.5 dex. In these simulations (see Figure 3 in YA18), the
stellar material containing a standard chemical composition at the
stellar convective envelope goes down to be processed in the internal
radiative zone. There, the original $^{7}$Be element increases its abundance
up to a maximum due to the constructive nuclear process $^{3}$He ($^{4}$He,$\gamma$)
$^{7}$Be. According to YA18, this downward action requires a processing
time of 423 yr.

In this internal region with very high temperatures, the $^{7}$Li
abundance is very low due to the destructive reactions $^{7}$Li (p,$\gamma$) $^{8}$Be and
$^{7}$Li (p,$\alpha$) $^{4}$He. Here, $^{7}$Be attains its maximum abundance and begins to
increase the $^{7}$Li abundance by means of $^{7}$Be (e$^{-}$, $\gamma$) $^{7}$Li. Then, the upward
conveyor requires a fast motion with a high velocity to bring this
material, where $^{7}$Li is the dominant element, to the external envelope.
This fast motion is necessary to attain cooler regions and avoid the
destruction of Li$^{7}$ by the mentioned proton processes. YA18 obtains this
super Li-rich giant star in an upward time of only 47 yr. During that time,
$^{7}$Be suffers a moderate abundance decrease (see figure 3 in YA18).

Unfortunately, YA18 does not provide this $^{7}$Li velocity, but we can use
this time interval of 47 yr to obtain an approximate value for this velocity that
can be compared to the necessary upward velocity to transport other
elements such as C and N. For these last elements,  \cite{Eggleton2008} gave values around 2 cm/s; this is sufficient to mix a stable
region with the outer convective envelope in times much shorter than
the evolution times given by these authors. By using similar
atmospheric scales of this last reference as equal to 10\,R$_\odot$, we obtain a
velocity of $\sim$\,470\,cm/s for this distance with a time of 47 yr for the
outward transport of $^{7}$Li. This last velocity is much larger than that
required for C and N. We can conclude that due to this very large
difference in velocities, no correlation is expected between the $^{12}$C/$^{13}$C
ratio and Li. Concerning this correlation, we can refer to the first study that considered field Li-rich giant stars and their corresponding $^{12}$C/$^{13}$C ratios, where a complete absence of correlation was found by \cite{daSilva1995}.
Later, other studies on field
giant stars by \cite{AguileraGomez2023} showed that a correlation
between these carbon ratios and Fe exists, but no correlation with Li is
evident. Additionally, no clear correlations between these two
parameters ($^{12}$C/$^{13}$C and Li) appear to exist among giant stars in very different
environments, such as those in open clusters \citep{Gilroy1989} and
among super metal-deficient stars\citep{Molaro2023}.

\section{ The lithium shell model}

\begin{figure}
    \centering
    \includegraphics[width=.99 \hsize]{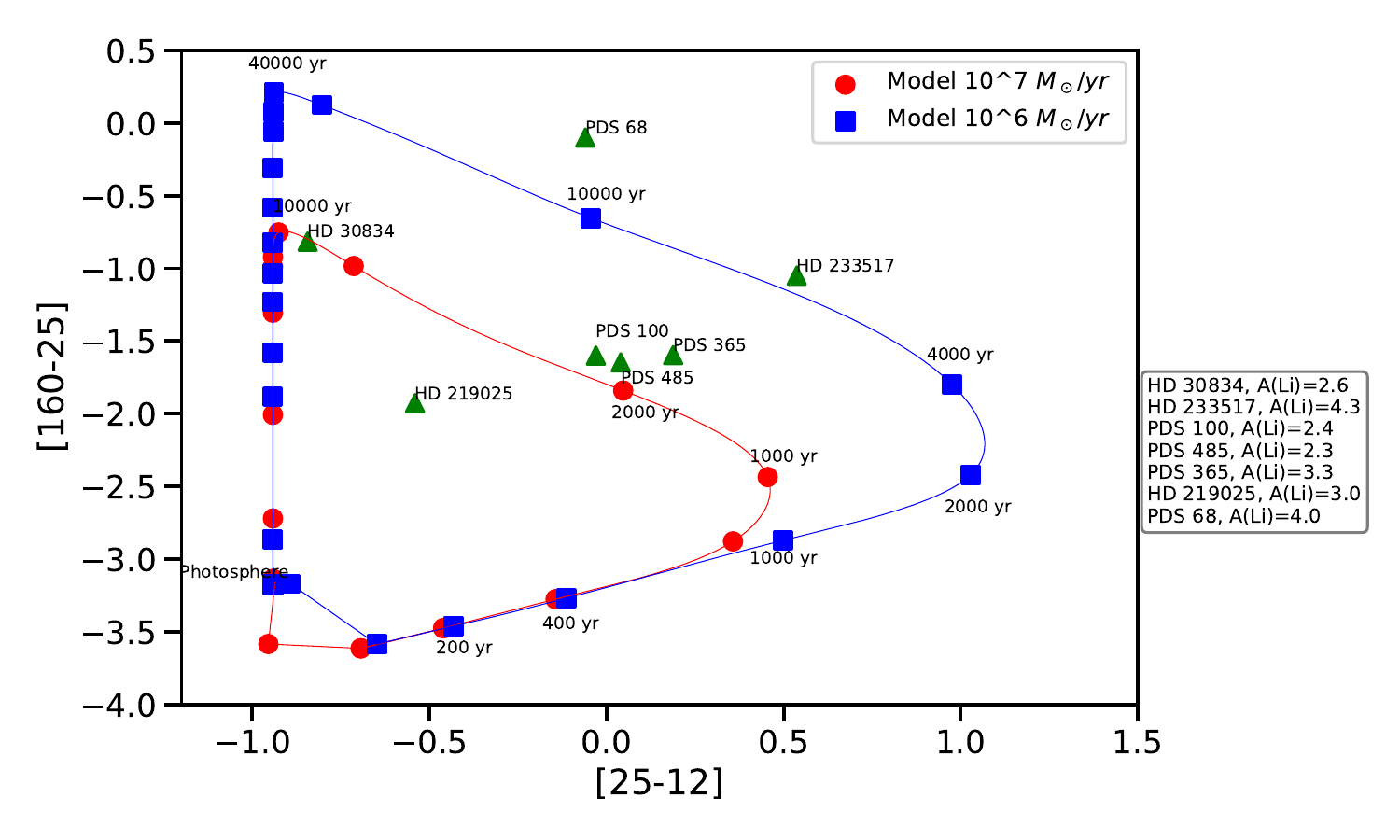}
    \caption{Model of stellar shell ejections represented here in a
colour-colour diagram with wavelengths at 12\,$\mu$m, 25\,$\mu$m, and 160\,$\mu$m,
with two loops indicating the growing in time of the shells (see text).
The red trajectory is for stellar mass losses of 10$^{-7}$\,$M_\odot$/yr and the blue is
for 10$^{-6}$\,$M_\odot$/yr. Green filled triangles indicate observed
colours of some giant stars together with their respective Li abundances.
Both axes are represented by [$\lambda_1$ - $\lambda_2$] = log ($\lambda_2F_1$) - log ($\lambda_1F_2$). The
observed fluxes (F)in Jy were obtained from \cite{Rebull2015} and
AKARI and HERSCHEL catalogues.
       }
     \label{fig5}
\end{figure}

Our proposed general scenario for the Li enrichment in low-mass giant
stars predicts the ejection of stellar shells characterising quite strong
stellar winds, as mentioned before. These shells are the observed IR
excesses associated with the Li-rich stars. Initially, these IR sources
were discovered and identified using the Infrared Astronomy Satellite
(IRAS). Later, \cite{Rebull2015} revisited all these IR sources of Li-rich
giant stars by means of the WISE explorer. In the original model  \citep{delaReza1996,delaReza1997},
considering IRAS sources, the shells were
immediately detached from the star, ejected, and lost in the interstellar
medium. This situation will probably lead to a sort of quasi-explosive
situation resembling the Li-flash thermal instability as suggested by
\cite{Palacios2006}. However, in \cite{delaReza2012}, a more
plausible conceptual physical modification was made without
changing the structure of the model. In this new conception, the shells
continue to be attached to the star until the completion of the internal Li-enrichment mechanism. Once this enrichment process is finished, the
shell detaches from the star and is finally ejected into the interstellar
medium. This new detachment concept has the advantage that, contrary
to the first model, the size of the shell can be estimated. In any case, the
increase of the shell follows an evolution directly related to a $V_s = R/t$
mode, in which $V_s$ is the velocity of the evolution of the size of the shell and
$R$ is the size of the shell. The time t is related to the lifetime of the shell.
When the shell detaches from the star, R then measures the size of the
shell plus the distance to the star. In a colour-colour diagram (Fig. \ref{fig5}),
constructed with IR colours at 12\,$\mu$m, 25\,$\mu$m, and 160\,$\mu$m, we show the
evolution paths of the shells. They begin in a box where the giant stars
present no IR excesses, which is characterised by normal Li-poor giant
stars. As time increases, the shell evolution forms a loop, finally
returning to the initial box. Following \cite{delaReza1996,delaReza1997} and \cite{delaReza2012}, the sizes of the loops depend mainly on the
stellar parameters $V_s$ and stellar mass loss. To a lesser extent, they depend on
the stellar radius and the effective temperature.

The model requires low evolution velocities of the shell's expansion, of
the order of $\sim$\,2\,km/s, to cover the positions of the observed fluxes at
those wavelengths. Mean values of the stellar radii of 20\,R$_\odot$ and of the
$T_{\rm eff}$ temperatures of 4900\,K have been adopted. The shell evolution 
curves contain the seven Li-rich giant stars with strong IR excesses
shown with green triangles in Fig. \ref{fig5}; these are HD\,233517, PDS\,68,
PDS\, 100, PDS \,365, PDS\,485, HD\,219025, and HD\,30834. The Li abundances for each object are mentioned in
the same figure. As can be seen in this figure, the stellar episodical
mass losses appear dependent on the Li abundances, with very high
mass losses of $\sim$ 10$^{-6}$\,M$_\odot$/yr for stars with Li abundances equal to or
larger than 4.0\,dex. Stars with Li abundances between 2.3\,dex and 3.3\,dex require slightly lower mass losses of the order of 10$^{-7}$\,M$_\odot$/yr.
It is interesting to note that larger mass losses, meaning more
energetic processes, appear necessary to produce larger quantities of
Li.
 \subsection{Considering whether lithium-enriching giant stars are the only ones
with IR excesses}

In this work, one of the most important points consists of affirming that
only giant stars that are in the process of being enriched with new
lithium present IR excesses. This means that normal lithium-poor giants
do not show these kinds of IR excesses. The validity of this last strong
condition requires a detailed examination of the different surveys to
detect IR excesses in giant stars. The majority of the different censuses
were completed using the IRAS database, which has been replaced by
the much more efficient WISE database. The surveys looking for IR
excesses in giant stars were made with very different star samples.

First, \cite{Jura1990}, collecting 100 bright G and K giants, noted the
absence of IR excesses in G giants and finally selected only two K
giants with low fluxes at 60 $\mu$m IRAS. \cite{Zuckerman1995}
considered 40,000 giants and found fewer than 300 possible excess
giants. For this last survey, we examined their published list of stars and
found, among K-type giants, only two objects with significant IRAS
excesses. These are HD\,19745 and HD\,219025, which are known to be
very lithium-rich stars \citep{delarezadasilva1995}. Both giants have
strong WISE excesses measured by W1 -- W4, corresponding to values
of 2.3 and 3.7. There is a third known lithium-rich giant with a lithium
abundance of $\sim$ 1.9 (HD\,30834), measured with lower quality IRAS data,
for which a WISE excess cannot be estimated. All the other giant stars
on this list have, when WISE measurements are possible, W1--W4
values less than or equal to 0.1, signifying no significant excesses.

Later, \cite{Plets1997} explored a large number of G, K, and M-type
giant stars, detecting among the K giants only two Li-rich stars, namely
HD\,30834 and HD\,146850. We searched for all K giants listed in its
appendix for which Li abundances are not provided, and we did not find
any star with a WISE W1--W4 value larger than 0.1, confirming the
absence of real IR-excess giant stars.
Later, a census with a similar purpose appeared in \cite{Kumar2015}
with a sample of 2000 low-mass K giants. Their results can be divided
into two parts. First, they published a list in their Table 1 of 40 lithium-rich giants, which also includes additional giant stars from the literature.

From this list, only seven giants, which are those taken from the
literature, have very strong IR excesses with WISE values between 2.3
and 7.3. All of the other lithium-rich stars have WISE values less than
or equal to 0.1 and are practically without IR excesses. Secondly, they
mention another list containing about two dozen K giants with
detectable far-IR excess, and, surprisingly, none of them are lithium-rich
giants'. Unfortunately, they do not publish this list, and we have no
means to examine it to know which stars they are, their precise number,
and what their actual degree of far-IR detectability' is. In an
attempt to understand which stars belong to this unknown list, we
explored, as far as possible, their three mentioned sources of data.
These are \cite{Kumar2011}, \cite{Liu2014}, and \cite{Adamow2014}.

Only the last two references present published lists containing lithium-poor giants. We considered all of the tables (Table 1 for K giants only) in
\cite{Liu2014} and Table 1 of \cite{Adamow2014}. As a result, we did
not find any evidence of lithium-poor normal giants with IR excesses
larger than W1--W4 $=$ 0.1.

\cite{Kumar2015}, which claimed that the physical coincidences of lithium-rich giants with IR excesses are very rare, are mainly based on a
misunderstanding that frequently occurs in the literature, and our work
here attempts to dissipate it. This misunderstanding is that lithium-rich
giants have a very low probability of having IR excesses because, as
seen here, especially in Sect. 4.1, their lithium-rich lifetimes are of the
order of 10$^{7}$ years, which is much longer than the estimated lifetimes
of the lithium short-enrichment episode; this is of the order of 10$^{4}$
years, and during this period the giant star presents an IR excess. In conclusion,
we can say that, up to this moment, there appear to be no lithium-poor
normal giants presenting IR excesses, at least with our adopted WISE
criteria of W1--W4 larger than or equal to 0.5 considered in our work
(Table 1) as a real excess. It thus appears that only Li-enriching stars,
meaning giant stars in the process of being enriched with fresh Li, are
the only ones presenting IR excesses. Future larger volumes of data
catalogues of giant stars merging with the WISE database could provide a
definitive answer on these matters.

\subsection{ Lithium stellar lifetimes}

The model does not specify the moment when the shell is detached
from the star. In any case, the loops close in a few million years when
the shell is expected to be completely lost. What is important, however, is that
the shells are expected to have short lifetimes. One way to approach a
somewhat more realistic lifetime consists of considering an age when
the time of the trajectory in the loop coincides with the observed
position of the star flux. As seen in Fig. \ref{fig5}, ages are around 2000\,yrs
for stars with Li abundances between 2.3 and 3.1\,dex, and between 
$\sim$\,5000\,yrs and $\sim$\,10000\,yrs for stars with larger Li abundances. In this
way, we are measuring the time of the fresh Li enrichment in these giant
stars. In the same way, we are characterising the lifetimes of this short
event.
There is another approximate way (not dependent on the preceding
model, but more general) to evaluate the lifetimes of Li in giant low-mass
stars. This involves simple considerations of the total number of stars
studied here. If we take into account the total number of explored giant
stars in the LAMOST survey (GA19), which is 814268 stars, the resulting
Li-rich giant stars are 10535 with Li abundances equal to or larger than
1.5\,dex. This means a fraction of 0.0129. In this work, we detected 421
RC stars and 196 RGB stars presenting IR excesses, or a total of 617
stars. This total number represents a fraction of 0.058, or almost 6\%, of
the Li-rich stars in the LAMOST catalogue. If we consider crude values of
the lifetimes of stars in the RGB stage as 10$^{9}$\,yrs and 10$^{8}$\,yrs for RC stars,
or 1.1\,x\,10$^{9}$\,yrs, we obtain, considering fractions as indicators of lifetimes,
that the Galactic Li lifetime is approximately 0.0129\,x\,1.1\,x\,10$^{9}$\,=\,1.4\,x\,10$^{7}$\,yrs. Now, the fraction of giant stars with IR excesses or losing mass will
be approximately 0.058\,x\,1.4\,x\,10$^{7}$ $=$ 8\,x\,10$^{5}$\,yr. We note that this value is
contained in the total time of the Li enrichment loops of Fig. \ref{fig5}. We
saw, however, that in those loops, more realistic lifetimes of the prompt
Li enrichment could be one or more orders of magnitude smaller. What we learn
here is that if 6\% of the stars are being enriched with Li, the remaining
94\% of Li-rich stars maintain their new Li for longer periods of $\sim$\,10$^{7}$\,yr.
We note, moreover, that this is a general evaluation of the involved future Li-star depletion. In RC giant stars, which represent a different physical
situation than RGB stars, much shorter Li depletion times could exist.

\subsection{Chromospheric activation and the lithium properties}

\begin{figure}
    \centering
    \includegraphics[width=.99 \hsize]{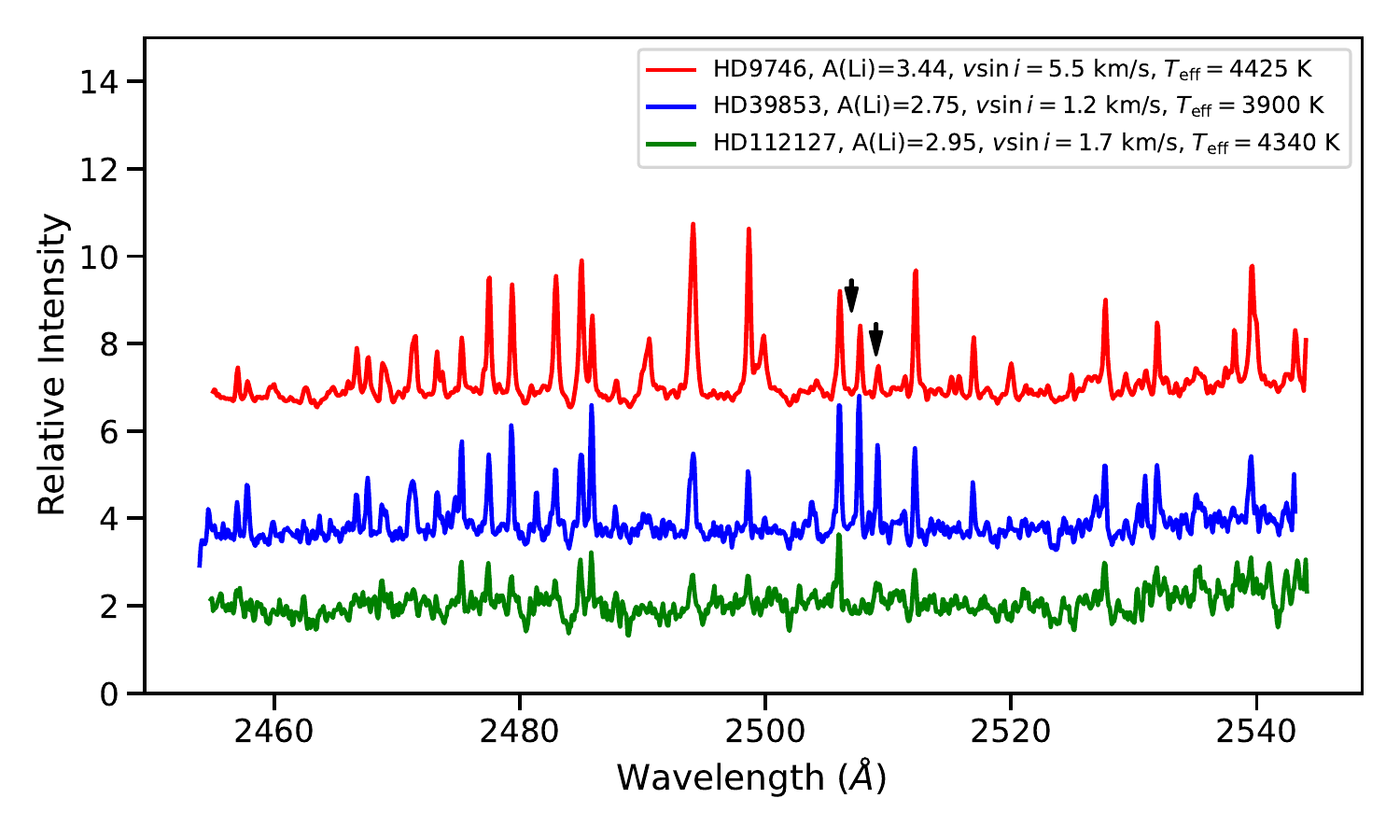}
    \caption{UV \textit{Hubble} spectra of three Li-rich giant stars --HD\,9746,\,HD
39853, and HD\,112127-- are shown here \citep[See also][]{Drake2018} with
their respective Li abundances and $v\,\sin{i}$ values. Emission lines are
mostly due to Fe{\rm II}. The place of the faint neutral boron resonance lines
is indicated by arrows.
}
\label{fig6}
\end{figure}

Using the \textit{Hubble} Space Telescope, we observed the spectra of the four
Li-rich giant stars (HD\,9746, HD\,39853, HD\,112127, and HD\,787) between
$\sim$\,2460\,\text{\AA} and $\sim$\,2540\,\text{\AA} \citep{Drake2018}. The original objective was to measure the two neutral boron lines at 2496.771\,\text{\AA} and 2497.725\,\text{\AA} to
test the engulfing-planet scenario, as proposed to explain the origin of
Li-rich giant stars.
To our surprise, we found that the intensity of both
the continuum and the spectral UV lines increased apparently according
to the growing Li abundances. However, this relation with Li is indirect,
via the chromospheric excitation. This is shown in Fig. \ref{fig6}, where the
UV spectra of stars HD\,9746, HD\,39853, and HD\,112127 are presented in
order. Their corresponding important stellar parameters, including Li
abundance, $T_{\rm eff}$, and rotational $v\,\sin{i}$ velocity, are as follows: for 
HD\,9746: 3.44, 4425\,K, and 5.5\,km/s; for HD\,39853: 2.75, 3900\,K, and 1.2\,km/s; and for HD\,112127: 2.95, 4340\,K, and 1.7\,km/s. All these parameters
are taken from \cite{Jorissen2020}, with the exception of the HD\,112127 $v\,\sin{i}$ value  taken from \cite{deMedeiros1996}. 
We note that for HD\,9746,
\cite{deMedeiros1996} proposed an even larger $v\,\sin{i}$ value of 8.7\,km/s. It
thus appears that the main parameter controlling the intensities of the
UV lines is that of the rotational velocity. \cite{Jorissen2020} considered
$v\,\sin{i}$ values larger than or equal to 5.0\,km/s as those of rapid rotators.

\begin{figure}
    \centering
    \includegraphics[width=.99 \hsize]{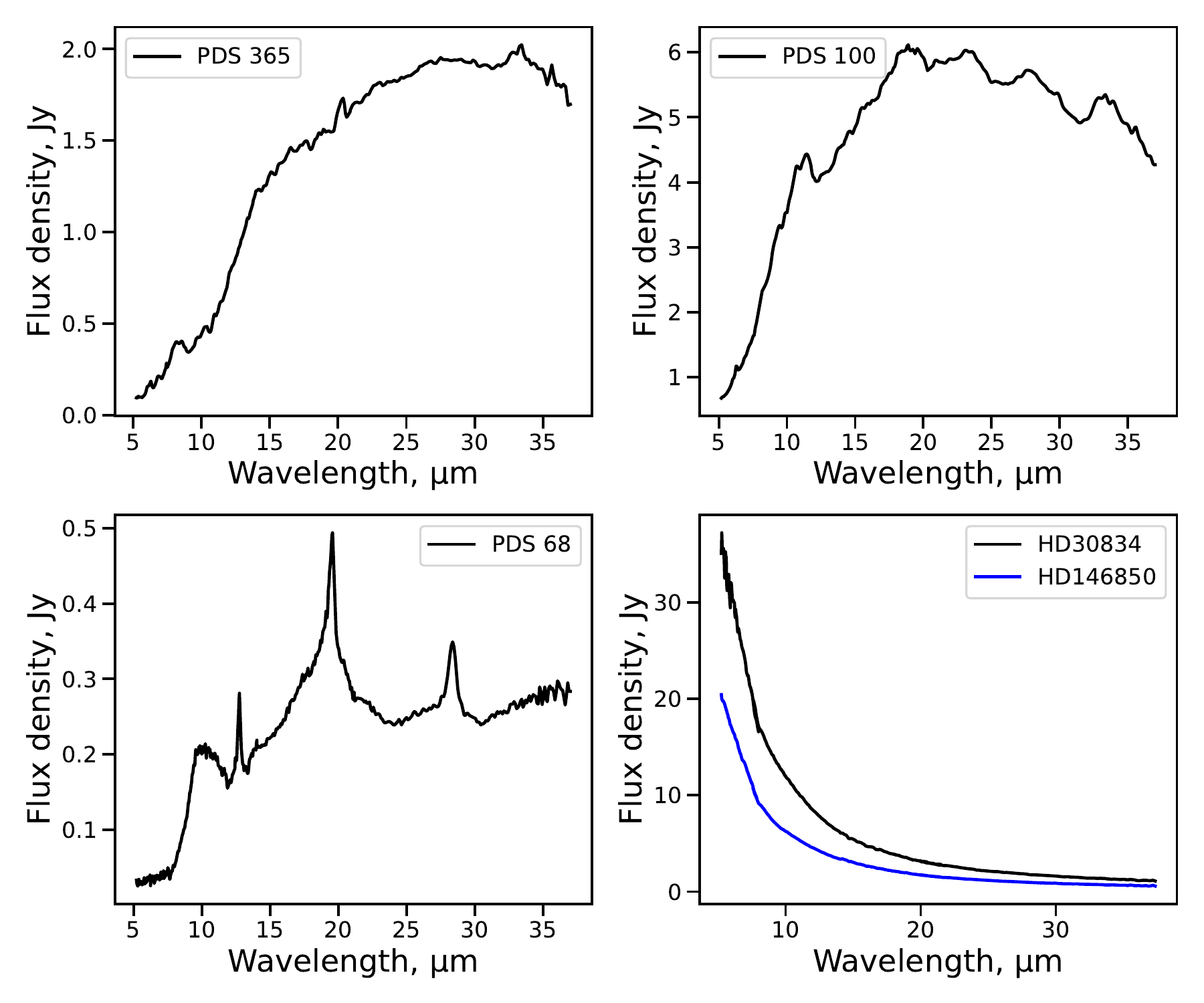}
    \caption{NIR spectra obtained with Spitzer telescope are shown
here for some Li-rich and super Li-rich giant stars: a) PDS\,365, A(Li)\,$=$\,3.3
dex; b) PDS 100, A(Li)\,$=$\,2.4; c) PDS\,68, A(Li)\,$=$\,4.0; and d) HD 30834, A(Li)\,$=$\,2.6 and HD\,146850, A(Li)\,$=$\,1.6. The first three stars present a strong emission continuum
due to their attached shells, and following our model they are in the
sudden Li-enriching stage. In the last two stars, the Li-enrichment
process is finished and the shell, which is no longer present, is considered to be
ejected, whereas the star keeps its Li for a certain time. Superposed on
these emission continuums are localised emission peak features due to
the presence of complex aromatic organic and inorganic structures (see
DLR15).
}
\label{fig7}
\end{figure}

The involved emission lines, as shown in Fig. \ref{fig6}, are mainly Fe{\rm II} lines
\citep{Judge1991}. Among these lines, for
example, the 2507\,A and 2509\,A lines are fluorescence lines pumped by
H Ly$\alpha$ \cite{Johansson1993}, pointing to the existence of very
strong H Ly$\alpha$ emission. What is important in this picture is that the
chromospheres of these stars appear to be activated. For a correlation
between activity and high Li abundance, we suggest the works of \cite{Fekel1993} and \cite{Drake2002}. Another recent study concerning chromospheric and Li relations, this time involving the He I
line at 10830\,\text{\AA}, is that of \cite{Sneden2022}. Here, the very Li-rich
single giant star HD 233517 \citep{Strassmeier2015,Fekel1996}, with a Li abundance of 4.3\,dex, received special attention. This star has by far the strongest He\,I\,10800\,\text{\AA} and is
separated from all the other Li-rich and Li-poor giant stars in the
mentioned study. In this aspect, searching for a tighter chromospheric--Li abundance relation, it will be of particular interest to measure the
intensity of the He I line at 10830\,\text{\AA} in other very Li-rich and Li-rich giant
stars (such as those presented as in Fig. \ref{fig7} of the next subsection)
that have a strong mass loss, such as HD\,233517. In the qualitative general
scenario presented in this work, the surface rotation of giant stars
depends on how the AM transfer is realised. It also depends on the
internal rapid upward mass transfer that produces the surface Li
enrichment. We can propose that more energetic mass losses will
produce more chromospheric activation during their journey from the
stellar interiors to the external wind regions. Larger abundances of
lithium will then appear to be present mainly in rapid rotators.

 Even if the majority of low-mass giant stars appear to be quiet and inactive stars, there is a fraction of them presenting significant  UV-emission excess, which shows  a good correlation with stellar rotation
velocities \citep{Dixon2020}. This rotation-activity relation, which has
been largely studied for dwarf stars, applies, according to these authors, to
RGB and RC giants.

In this context, and regarding   Li-rich giants, it is
surprising to see here in Fig. \ref{fig6} that these few giant stars 
appear to correlate between UV line intensities and $v\,\sin{i}$.
 Even more surprising is that the measured   $v\,\sin{i}$ value of the
highest intensity UV-emission spectrum star (HD\,9746) is 8.7\,km/s \citep{deMedeiros1996} and
not 5.5\,km/s as shown in the figure. 

Regarding Li giants, we propose, as
previously mentioned, an extension of the study of chromospheric
activity in Li-rich stars by \cite{Sneden2022} using the chromospheric
He line at 10830\,\text{\AA};  to this study, we would add rapidly rotating giant stars of varying Li abundances and IR excesses. This can provide
more information to examine whether stellar rotation is, in fact, the
dominant driver of activity.

Mass-loss rate calculations are also being studied in giant stars by 
a different approach than the dust-based mass-loss calculations based on IR excesses used here
\citep{delaReza1996}. These are based on the
nonlinear propagation of magnetic-origin Alfvén waves 
\citep[see among others, ][]{Cranmer2011,Airapetian2010}. 
Although discussing these models is beyond the scope of this work, we can mention that they were aimed at reproducing stellar mass-loss values of normal Li-poor giants.

This is the case, for example, of the known K giant star Alpha
Tau in \cite{Airapetian2010}, with a mass loss of 1.6x10$^{-11}$\,M$_\odot$/yr. In
these conditions, it will be a challenge for these models to reproduce
the very brief mass-loss episodes  of 10$^{-7}$--10$^{-6}$\,M$_\odot$/yr, as seen in Fig. 5,
which are necessary to  explain the rapid Li enrichment episodes proposed in our scenario.

\subsection{ Complex organic and inorganic particles in the winds
of very lithium-rich giant stars}

The first aromatic organic material (PAH) was detected in the well-known Li-rich giant star HD\,233517 \citep{Jura2006}. The observations
were realised in the NIR (5\,$\mu$m -- 38\,$\mu$m) spectra by means of the Spitzer
telescope. \cite{Jura2006} proposed that this material originated in a
hypothetical disc around this star. Later,  DLR15
 detected complex organic compounds and the
presence of inorganic particles in several other Li-rich giant stars.
These results indicate that this material is originated in shells around
these stars. In DLR15, it was proposed that the observed organic
aromatic and aliphatic compounds were formed in the winds by
episodical mass losses of the order of 10$^{-7}$\,M$_\odot$/yr to 10$^{-6}$\,M$_\odot$/yr.

Additionally, the inorganic particles resulted from a stellar wind
dragging of the debris disc that remained since these were A-type
stars on the main sequence. These discs were probably partially or
completely eroded by the action of these strong winds. Among the
several inorganic particles detected were Forsterite, enstatite, and
several other compounds. The main characteristic of the mentioned NIR
spectra is the presence of an important continuum emission in which
narrow emission features are superposed at very specific wavelengths.
The presence of these features enables us to identify the nature of the
organic and inorganic material under consideration (see DLR15). As an
example, some of these spectra are shown in Fig. \ref{fig7} corresponding,
respectively, to the giant Li-rich stars PDS\,365, PDS\,100, and PDS\,68.
These stars are being enriched with Li and are consequently forming a
shell represented by strong NRI continuum emission spectra. For
illustration and comparison purposes, we show, in the same Fig. \ref{fig7},
two giant stars: HD\,30834, a Li-rich star with a Li abundance of $\sim$\,2.6\,dex
\citep{Takeda2017}; and a moderately Li-rich star, HD\,146850, with a Li
abundance of 1.6\,dex \citep{Castilho1999}.

Unlike the PDS stars in the figure, these two stars have no shells
detected in the NIR between 5\,$\mu$m and 38\,$\mu$m observed by SPITZER;
however, HD 30835 has an important observed flux at 160\,$\mu$m (see
Fig.\,\ref{fig5}) suggesting that its shell is detached and presently located at a
larger distance from the star. Sadly there are no FIR data for HD
146850. Following our scenario, HD\,30834 and HD\,146850 have
finished their Li-enrichment process but maintain their new Li.

\subsection{ Lithium properties of red-clump giant stars}

Since the first detection of a Li-rich RC star made with the help of
asteroseismology by \cite{Silva-Aguirre2014}, a large contribution to
this subject has recently appeared in the literature. We limit
ourselves to mentioning these references without writing about them further,
except for those relevant to the following discussion concerning our
proposed scenario. In chronological order, we mention, among others,
the following references: \cite{Kirby2016,Kumar2020,Schwab2020,Martell2021,Mori2021, Deepak2021,Magrini2021,Yan2021,Singh2021,Zhou2022,Chaname2022,Mallick2022,Sayeed2024}. A recent work by \cite{Susmitha2024} opens a new avenue to study the effect of metallicity on the Li production in RC giant stars.

In this work, we present new data on 421 RC Li-rich stars with
IR excesses. All of these stars cover a wide range of Li abundances,
from 1.5\,dex up to almost 5\,dex, thus presenting important mass-loss
properties.
Recently, \cite{Mallick2022} conducted the first study of RC stars
considering IR excesses, confirming that the circumstellar shells lost by
these stars have short lifetimes of the order of 3000 yr or less, as we
mentioned before for all Li-rich giant stars. It must be noted that no
work using the He-flashes existing in this evolutionary stage has been
able to explain the super Li-rich state of these stars. If our general Li
scenario is correct, these stars are being enriched with fresh Li due to
their observed IR excesses. We suggest that these same stars may be
at least partially forming new Li through He flashes. By adding these
two separate Li contributions, we may be able to explain the presence
of the observed super-Li-rich RC giant stars.

\subsection{ Lithium abundance and stellar rotation}

There may be a relation between the stellar Li abundance of giant stars and
their rotational stellar velocity. This kind of problem is largely
mentioned in the literature and is left unanswered. The
probable reason for this is the difficult physical mechanisms
involved. As an example, we can take the giant stars considered in this work for which we have
measurements of the Li abundances and projected rotational
velocities. Even though they are not numerous, they can provide a first
insight into this question.

\begin{figure}
    \centering
    \includegraphics[width=.7 \hsize]{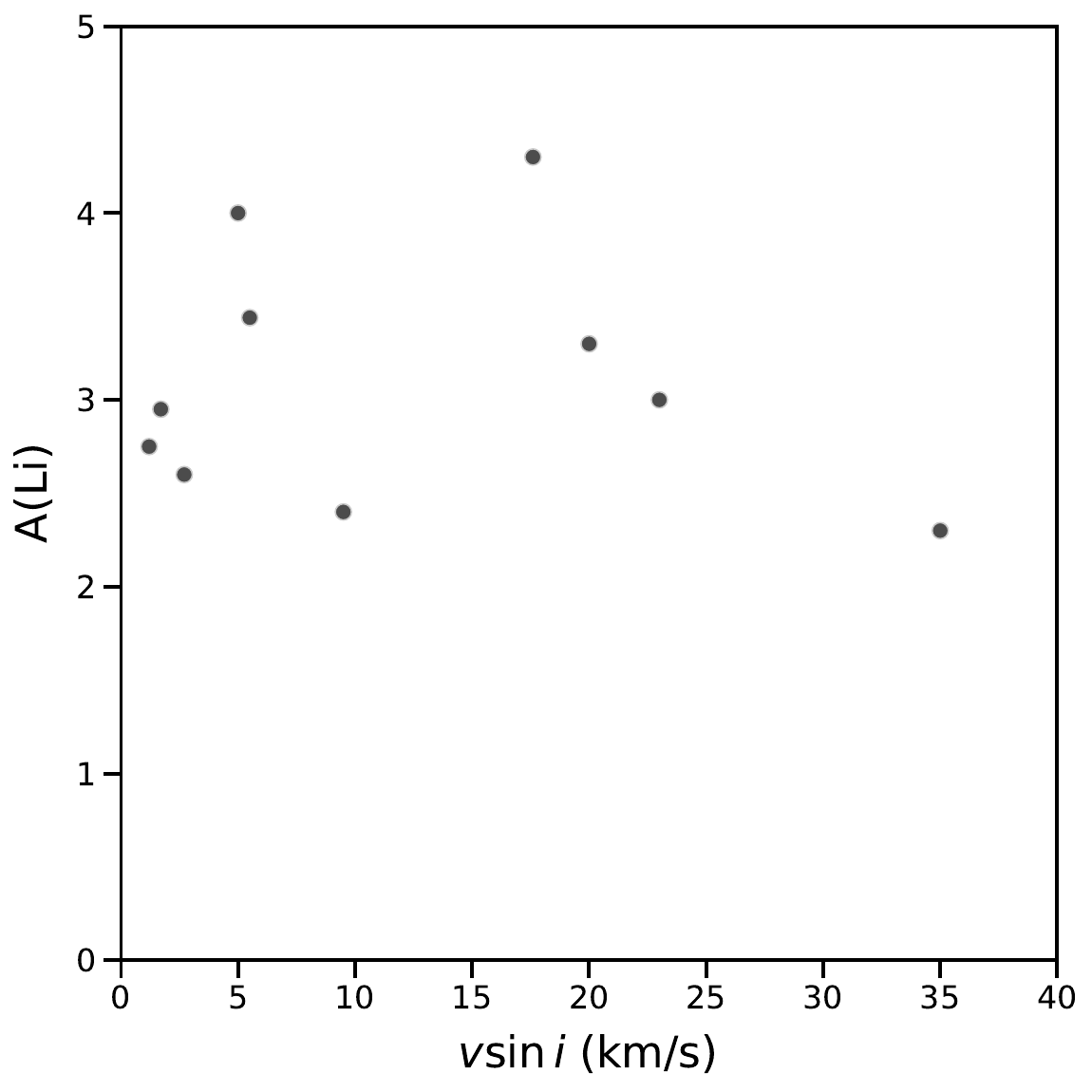}
    \caption{ Projected velocity $v\,\sin{i}$ as function of Li abundance for all Li-rich giant stars discussed in this work. }
\label{fig8}
\end{figure}

These stars have been discussed in various
sections of this work and are the following, listed in order of name, Li
abundance, and $v\,\sin{i}$ in km/s: HD\,9746, (3.44, 5.5; HD\,30834, (2.6, 2.7); HD\,39853, (2.75, 1.2); HD\,112127, (2.95, 1.7); HD\,219025, (3.0, 23); HD\,233517, (4.3, 17.6); PDS\,68, (4.0, 5.0); PDS\,100, (2.40, 9.5); PDS\,365, (3.3, 20); PDS\,485,
(2.3, 35.0). The references of velocities for the star PDS\,100 can be found in
\cite{Reddy2002}, and for stars PDS\,68 and PDS\,485 they can be found in \cite{Reddy2005}.
Only three of these stars do not present IR excesses. A simple
inspection of these numbers shows that a linear relation between the Li
abundance and $v\,\sin{i}$ is inexistent, even if all these stars have Li
abundances larger than 2.0\,dex, and these stars are rapid rotators if we
consider rotational velocities larger than 5.0 km/s as rapid rotators, as
mentioned in Sect. 4.3. This discussed non-correlation between
the Li abundance and $v\,\sin{i}$ can be seen in Fig. \ref{fig8}.

Concerning the physical aspects behind this
problem, we can first mention, as also presented in Sect. 4.3,
that it is mainly the activation of the chromosphere (in our scenario by a
mass transfer coming from the stellar interior and crossing the
chromosphere) that probably increases the stellar rotation as observed (see Fig. \ref{fig6}). Moreover, this activation is not directly related  to the $^{7}$Li
transport to the external stellar layers.

Another aspect, and maybe the
most difficult, is that the rotation of the external stellar envelope of giant
stars depends on how the internal transfer of the stellar AM is actually
realised, and this is not well known. In conclusion, considering all these
problems, it can be natural not to find a one-to-one relation between the
Li abundance and rotation. Nevertheless, as mentioned in Sect. 4.3
 a general and crude relation exists between rapid stellar rotators
and high Li abundances.

\section{ Discussion and conclusion}

Working solely with the LAMOST DR7 LRS catalogue (GA19),
which contains 10,535 giant stars with Li abundances ranging from 1.5
dex to 4.9\,dex, and the WISE catalogue, we detected 617 Li-rich stars
exhibiting IR excesses. To differentiate between RGB and RC giant stars
with IR excesses, we employed a classification based on log-g values.
This selection method aligns with that used by YA21 for Li-
rich stars without IR excesses, based on asteroseismology. Our
findings demonstrate that the distributions of RGB and RC stars from
\cite{Yan2021} and ourselves are comparable. 

In this way, we were able to find 421 RC stars and 196 RGB stars
with IR excesses, as displayed in Fig. \ref{fig3}, for Li abundances ranging
from 1.5\,dex up to 4.8\,dex. However, their distributions are different.
RGB stars have a general maximum Li abundance aof round 2.6\,dex,
whereas RC stars attain maximum values of 4.8\,dex. In other words, RC
stars are by far the most Li-rich stars.

The distribution of our general qualitative scenario for Li enrichment,
which is valid for all single low-mass giant stars, proposes that a
sudden Li enrichment occurs, accompanied by an also sudden
formation and ejection of circumstellar material. This scenario,
described in detail in Sect. 3, includes one speculative key
ingredient: the existence of an unknown internal sporadic instability,
likely of magnetic origin, that can rapidly transport mass from the upper
H-burning zone upwards. This process is part of the unknown evolution
of total angular momentum. It is important to note that the key
ingredient to form Li-rich and super Li-rich giant stars consists of
rapidly transporting the internal material, where $^{7}$Li is the dominant
element, to the external convective envelope. Based on nuclear
simulations found in the literature, we estimate the upward velocity of
the $^{7}$Li material to be $\sim$\,500\,cm/s, a much larger value than that required
to transport other elements such as C, which is of the order of 2\,cm/s. Due
to this significant difference in velocities, an anti-correlation between
the $^{12}$C/$^{13}$C ratio and Li is expected and is actually observed.

The expected sudden enrichment of Li produces the formation and
ejection of circumstellar shells. By applying a shell-model evolution to
Li-rich giant stars with large IR excesses from the literature, we find quite short lifetimes for the formation and subsequent ejection of
these shells. These lifetimes are on the order of 2000 years and 10000
years, thus characterising the expected rapid Li-enrichment process.
The stellar mass-losses involved in this process are very short and
several orders of magnitude larger than those of Li-poor giant stars. In
this model, the stellar mass losses appear to be proportional to the Li
abundances.

Based on the fraction of Li-rich giant stars with respect to Li-poor giant
stars in the LAMOST survey, which is 0.0129, and the crude values of
the lifetimes of the RGB and RC stages, respectively 10$^{9}$\,yrs and 10$^{8}$\,yrs, it
is found that the Li Galactic age is approximately 1.4\,x\,10$^{7}$\,yrs. Now,
considering our found fraction of stars presenting mass losses, which
is 0.058, we infer a total age of Li enrichment of 8 x 10$^{5}$ yrs. This time
coincides with the maximum age of the shell ejection for a complete
loop in the shell model. We note, however, that more realistic ages of the
shells are much shorter: on the order of 10$^{4}$ yrs or shorter.

It is expected that the internal mass coming from the stellar interior will
traverse the stellar chromospheres and activate them. By means of UV
spectra taken with the \textit{Hubble} telescope \citep{Drake2018} of three Li-
rich giant stars, with different Li abundances and different rotational
velocities, it is found that their stellar chromospheres appear more
activated as they have higher Li abundances. However, this activation
relation is more directed relatively to the rotational velocities, with the Li
abundances being a consequence of these activations. Chromospheric
studies by means of the 10800 He I line by \cite{Sneden2022} show
that the very Li-rich giant star HD\,233517, which is also a rapid rotator and
presents a very strong wind, has by far the strongest observed He I
line. 

To obtain a tighter chromospheric
relation between all these parameters, observing
other similar stars such as PDS\,68, PDS\,100, PDS\,365, and PDS\,485 with this He I line is recommended. In
fact, the majority of these same mentioned stars produce the formation
of complex organic and inorganic particles around the stars as
observed by the Spitzer Telescope (DLR15). If the organic material is
formed in their winds, these winds are sufficiently strong to partially or
totally destroy their remaining debris discs when these stars were main-sequence A-type stars, thus producing inorganic material in this way.

The study of the Li properties in RC stars is currently an active field of
research. In this work, we present new data on 421 RC Li-rich stars with
IR excesses. All of these stars cover a large range of Li abundances,
from 1.5\,dex up to almost 5\,dex, and therefore exhibit significant mass
losses. It must be noted that no previous work, utilising the He flashes
that occur in this evolutionary stage, has been able to explain the super-Li-rich state of these stars. If our general Li scenario is correct, these
stars are currently being enriched with fresh Li due to their observed IR
excesses. We suggest that if new Li is produced by means of He flashes
in these same stars, at least partially, we can infer that by combining
these two separate Li contributions, we may be able to explain the
presence of the observed super Li-rich RC giant stars.

Finally, in support of one of the main arguments of our scenario, in
which only giant stars that are in the process of being enriched with Li
present different degrees of IR excesses or mass loss, we carried out an
extensive search of surveys in the literature to determine if normal Li-poor giant stars could present IR excesses. To date, we find
no giant stars of this kind showing IR excesses such as those found in Li-enriching giants. 

We find that a linear relation between the Li stellar abundances and
the rotational velocities is inexistent and discuss the physical reasons
for this result in this paper.
In conclusion, if our scenario, which is based on a universal fast Li-enrichment-episode process that affects all stars, is correct, this can
explain why Li-rich giant stars are so few. This universal approach to
the Li problem was first proposed in \cite{delaReza1996,delaReza1997}. Later,
other works adopted this point of view, such as \cite{Kirby2012}, when
the existence of Li-rich giant stars in dwarf galaxies was discovered.
Additionally, recent works also adopt this universal approach. As
reported by \cite{Sayeed2024}, these are from the point of view of
observations by \cite{Kirby2016}, \cite{Kumar2020}, and \cite{Mallick2022}, and from expectations of theoretical models \citep{Schwab2020,Mori2021,Magrini2021}. In this respect, we quote the abstract words of \cite{Kirby2012}: `We consider the possibility that Li
enrichment is a universal phase of evolution that affects all stars, and it
seems rare only because it is brief'.

\section{Data availability}
Table 1 is only available at the CDS via anonymous ftp to \texttt{cdsarc.u-strasbg.fr} (130.79.128.5) or via \url{http://cdsweb.u-strasbg.fr/cgi-bin/qcat?J/A+A/}.

 \begin{acknowledgements}  
Several people have collaborated with me on various aspects during the
elaboration of this work, these are: Hong-Liang Yan, Félix Llorente de
Andrés, Carolina Chavero, Natalia A. Drake and Max Snoek. Many
thanks to all of them.  I also thank the anonymous referee for very useful comments.
\end{acknowledgements}

\bibliographystyle{aa} % style aa.bst
\bibliography{biblio} % your references biblio.bib

\begin{thebibliography}{83}
\expandafter\ifx\csname natexlab\endcsname\relax\def\natexlab#1{#1}\fi

\bibitem[{{Adam{\'o}w} {et~al.}(2014){Adam{\'o}w}, {Niedzielski}, {Villaver},
  {Wolszczan}, \& {Nowak}}]{Adamow2014}
{Adam{\'o}w}, M., {Niedzielski}, A., {Villaver}, E., {Wolszczan}, A., \&
  {Nowak}, G. 2014, \aap, 569, A55

\bibitem[{{Aerts} {et~al.}(2019){Aerts}, {Mathis}, \& {Rogers}}]{Aerts2019}
{Aerts}, C., {Mathis}, S., \& {Rogers}, T.~M. 2019, \araa, 57, 35

\bibitem[{{Aguilera-G{\'o}mez} {et~al.}(2016){Aguilera-G{\'o}mez},
  {Chanam{\'e}}, {Pinsonneault}, \& {Carlberg}}]{AguileraGomez2016}
{Aguilera-G{\'o}mez}, C., {Chanam{\'e}}, J., {Pinsonneault}, M.~H., \&
  {Carlberg}, J.~K. 2016, \apj, 829, 127

\bibitem[{{Aguilera-G{\'o}mez} {et~al.}(2023){Aguilera-G{\'o}mez}, {Jones}, \&
  {Chanam{\'e}}}]{AguileraGomez2023}
{Aguilera-G{\'o}mez}, C., {Jones}, M.~I., \& {Chanam{\'e}}, J. 2023, \aap, 670,
  A73

\bibitem[{{Aguilera-G{\'o}mez} {et~al.}(2022){Aguilera-G{\'o}mez}, {Monaco},
  {Mucciarelli}, {Salaris}, {Villanova}, \& {Pancino}}]{AguileraGomez2022}
{Aguilera-G{\'o}mez}, C., {Monaco}, L., {Mucciarelli}, A., {et~al.} 2022, \aap,
  657, A33

\bibitem[{{Airapetian} {et~al.}(2010){Airapetian}, {Carpenter}, \&
  {Ofman}}]{Airapetian2010}
{Airapetian}, V., {Carpenter}, K.~G., \& {Ofman}, L. 2010, \apj, 723, 1210

\bibitem[{{Bedding} {et~al.}(2011){Bedding}, {Mosser}, {Huber},
  {Montalb{\'a}n}, {Beck}, {Christensen-Dalsgaard}, {Elsworth}, {Garc{\'\i}a},
  {Miglio}, {Stello}, {White}, {De Ridder}, {Hekker}, {Aerts}, {Barban},
  {Belkacem}, {Broomhall}, {Brown}, {Buzasi}, {Carrier}, {Chaplin}, {di Mauro},
  {Dupret}, {Frandsen}, {Gilliland}, {Goupil}, {Jenkins}, {Kallinger},
  {Kawaler}, {Kjeldsen}, {Mathur}, {Noels}, {Silva Aguirre}, \&
  {Ventura}}]{Bedding2011}
{Bedding}, T.~R., {Mosser}, B., {Huber}, D., {et~al.} 2011, \nat, 471, 608

\bibitem[{{Behmard} {et~al.}(2023){Behmard}, {Sevilla}, \&
  {Fuller}}]{Behmard2023}
{Behmard}, A., {Sevilla}, J., \& {Fuller}, J. 2023, \mnras, 518, 5465

\bibitem[{{Cameron} \& {Fowler}(1971)}]{Cameron1971}
{Cameron}, A.~G.~W. \& {Fowler}, W.~A. 1971, \apj, 164, 111

\bibitem[{{Carlberg} {et~al.}(2012){Carlberg}, {Cunha}, {Smith}, \&
  {Majewski}}]{Carlberg2012}
{Carlberg}, J.~K., {Cunha}, K., {Smith}, V.~V., \& {Majewski}, S.~R. 2012,
  \apj, 757, 109

\bibitem[{{Casey} {et~al.}(2019){Casey}, {Ho}, {Ness}, {Hogg}, {Rix},
  {Angelou}, {Hekker}, {Tout}, {Lattanzio}, {Karakas}, {Woods}, {Price-Whelan},
  \& {Schlaufman}}]{Casey2019}
{Casey}, A.~R., {Ho}, A. Y.~Q., {Ness}, M., {et~al.} 2019, \apj, 880, 125

\bibitem[{{Casey} {et~al.}(2016){Casey}, {Ruchti}, {Masseron}, {Randich},
  {Gilmore}, {Lind}, {Kennedy}, {Koposov}, {Hourihane}, {Franciosini}, {Lewis},
  {Magrini}, {Morbidelli}, {Sacco}, {Worley}, {Feltzing}, {Jeffries},
  {Vallenari}, {Bensby}, {Bragaglia}, {Flaccomio}, {Francois}, {Korn},
  {Lanzafame}, {Pancino}, {Recio-Blanco}, {Smiljanic}, {Carraro}, {Costado},
  {Damiani}, {Donati}, {Frasca}, {Jofr{\'e}}, {Lardo}, {de Laverny}, {Monaco},
  {Prisinzano}, {Sbordone}, {Sousa}, {Tautvai{\v{s}}ien{\.{e}}}, {Zaggia},
  {Zwitter}, {Delgado Mena}, {Chorniy}, {Martell}, {Silva Aguirre}, {Miglio},
  {Chiappini}, {Montalban}, {Morel}, \& {Valentini}}]{Casey2016}
{Casey}, A.~R., {Ruchti}, G., {Masseron}, T., {et~al.} 2016, \mnras, 461, 3336

\bibitem[{{Castilho} {et~al.}(1999){Castilho}, {Spite}, {Barbuy}, {Spite}, {de
  Medeiros}, \& {Gregorio-Hetem}}]{Castilho1999}
{Castilho}, B.~V., {Spite}, F., {Barbuy}, B., {et~al.} 1999, \aap, 345, 249

\bibitem[{{Castro-Tapia} {et~al.}(2024){Castro-Tapia}, {Aguilera-G{\'o}mez}, \&
  {Chanam{\'e}}}]{CastroTapia2024}
{Castro-Tapia}, M., {Aguilera-G{\'o}mez}, C., \& {Chanam{\'e}}, J. 2024, \aap,
  690, A367

\bibitem[{{Chanam{\'e}} {et~al.}(2022){Chanam{\'e}}, {Pinsonneault},
  {Aguilera-G{\'o}mez}, \& {Zinn}}]{Chaname2022}
{Chanam{\'e}}, J., {Pinsonneault}, M.~H., {Aguilera-G{\'o}mez}, C., \& {Zinn},
  J.~C. 2022, \apj, 933, 58

\bibitem[{{Charbonnel} \& {Lagarde}(2010)}]{Charbonnel2010}
{Charbonnel}, C. \& {Lagarde}, N. 2010, \aap, 522, A10

\bibitem[{{Costa} {et~al.}(2002){Costa}, {da Silva}, {do Nascimento}, \& {De
  Medeiros}}]{Costa2002}
{Costa}, J.~M., {da Silva}, L., {do Nascimento}, J.~D., J., \& {De Medeiros},
  J.~R. 2002, \aap, 382, 1016

\bibitem[{{Cranmer} \& {Saar}(2011)}]{Cranmer2011}
{Cranmer}, S.~R. \& {Saar}, S.~H. 2011, \apj, 741, 54

\bibitem[{{da Silva} {et~al.}(1995){da Silva}, {de La Reza}, \&
  {Barbuy}}]{daSilva1995}
{da Silva}, L., {de La Reza}, R., \& {Barbuy}, B. 1995, \apjl, 448, L41

\bibitem[{{de la Reza} \& {da Silva}(1995)}]{delarezadasilva1995}
{de la Reza}, R. \& {da Silva}, L. 1995, \apj, 439, 917

\bibitem[{{de la Reza} \& {Drake}(2012)}]{delaReza2012}
{de la Reza}, R. \& {Drake}, N.~A. 2012, in Astronomical Society of the Pacific
  Conference Series, Vol. 464, Circumstellar Dynamics at High Resolution, ed.
  A.~C. {Carciofi} \& T.~{Rivinius}, 51

\bibitem[{{de la Reza} {et~al.}(1996){de la Reza}, {Drake}, \& {da
  Silva}}]{delaReza1996}
{de la Reza}, R., {Drake}, N.~A., \& {da Silva}, L. 1996, \apjl, 456, L115

\bibitem[{{de la Reza} {et~al.}(1997){de la Reza}, {Drake}, {da Silva},
  {Torres}, \& {Martin}}]{delaReza1997}
{de la Reza}, R., {Drake}, N.~A., {da Silva}, L., {Torres}, C.~A.~O., \&
  {Martin}, E.~L. 1997, \apjl, 482, L77

\bibitem[{{de la Reza} {et~al.}(2015){de la Reza}, {Drake}, {Oliveira}, \&
  {Rengaswamy}}]{delaReza2015}
{de la Reza}, R., {Drake}, N.~A., {Oliveira}, I., \& {Rengaswamy}, S. 2015,
  \apj, 806, 86

\bibitem[{{de Medeiros} {et~al.}(1996){de Medeiros}, {Melo}, \&
  {Mayor}}]{deMedeiros1996}
{de Medeiros}, J.~R., {Melo}, C.~H.~F., \& {Mayor}, M. 1996, \aap, 309, 465

\bibitem[{{Deepak} \& {Lambert}(2021)}]{Deepak2021}
{Deepak} \& {Lambert}, D.~L. 2021, \mnras, 505, 642

\bibitem[{{Denissenkov} {et~al.}(2024){Denissenkov}, {Blouin}, {Herwig},
  {Stott}, \& {Woodward}}]{Denissenkov2024}
{Denissenkov}, P.~A., {Blouin}, S., {Herwig}, F., {Stott}, J., \& {Woodward},
  P.~R. 2024, \mnras, 535, 1243

\bibitem[{{Dixon} {et~al.}(2020){Dixon}, {Tayar}, \& {Stassun}}]{Dixon2020}
{Dixon}, D., {Tayar}, J., \& {Stassun}, K.~G. 2020, \aj, 160, 12

\bibitem[{{Drake} {et~al.}(2002){Drake}, {de la Reza}, {da Silva}, \&
  {Lambert}}]{Drake2002}
{Drake}, N.~A., {de la Reza}, R., {da Silva}, L., \& {Lambert}, D.~L. 2002,
  \aj, 123, 2703

\bibitem[{{Drake} {et~al.}(2018){Drake}, {de La Reza}, {Smith}, \&
  {Cunha}}]{Drake2018}
{Drake}, N.~A., {de La Reza}, R., {Smith}, V.~V., \& {Cunha}, K. 2018, in
  Astrochemistry VII: Through the Cosmos from Galaxies to Planets, ed.
  M.~{Cunningham}, T.~{Millar}, \& Y.~{Aikawa}, Vol. 332, 237--241

\bibitem[{{Dumont}(2023)}]{Dumont2023}
{Dumont}, T. 2023, \aap, 677, A119

\bibitem[{{Eggleton} {et~al.}(2008){Eggleton}, {Dearborn}, \&
  {Lattanzio}}]{Eggleton2008}
{Eggleton}, P.~P., {Dearborn}, D. S.~P., \& {Lattanzio}, J.~C. 2008, \apj, 677,
  581

\bibitem[{{Fekel} \& {Balachandran}(1993)}]{Fekel1993}
{Fekel}, F.~C. \& {Balachandran}, S. 1993, \apj, 403, 708

\bibitem[{{Fekel} {et~al.}(1996){Fekel}, {Webb}, {White}, \&
  {Zuckerman}}]{Fekel1996}
{Fekel}, F.~C., {Webb}, R.~A., {White}, R.~J., \& {Zuckerman}, B. 1996, \apjl,
  462, L95

\bibitem[{{Gao} {et~al.}(2019){Gao}, {Shi}, {Yan}, {Yan}, {Xiang}, {Zhou},
  {Li}, \& {Zhao}}]{Gao2019}
{Gao}, Q., {Shi}, J.-R., {Yan}, H.-L., {et~al.} 2019, \apjs, 245, 33

\bibitem[{{Gilroy}(1989)}]{Gilroy1989}
{Gilroy}, K.~K. 1989, \apj, 347, 835

\bibitem[{{Holanda} {et~al.}(2020){Holanda}, {Drake}, \&
  {Pereira}}]{Holanda2020}
{Holanda}, N., {Drake}, N.~A., \& {Pereira}, C.~B. 2020, \aj, 159, 9

\bibitem[{{Johansson} \& {Hamann}(1993)}]{Johansson1993}
{Johansson}, S. \& {Hamann}, F.~W. 1993, Physica Scripta Volume T, 47, 157

\bibitem[{{Jorissen} {et~al.}(2020){Jorissen}, {Van Winckel}, {Siess},
  {Escorza}, {Pourbaix}, \& {Van Eck}}]{Jorissen2020}
{Jorissen}, A., {Van Winckel}, H., {Siess}, L., {et~al.} 2020, \aap, 639, A7

\bibitem[{{Judge} \& {Jordan}(1991)}]{Judge1991}
{Judge}, P.~G. \& {Jordan}, C. 1991, \apjs, 77, 75

\bibitem[{{Jura}(1990)}]{Jura1990}
{Jura}, M. 1990, \apj, 365, 317

\bibitem[{{Jura} {et~al.}(2006){Jura}, {Bohac}, {Sargent}, {Forrest}, {Green},
  {Watson}, {Sloan}, {Markwick-Kemper}, {Chen}, \& {Najita}}]{Jura2006}
{Jura}, M., {Bohac}, C.~J., {Sargent}, B., {et~al.} 2006, \apjl, 637, L45

\bibitem[{{Kirby} {et~al.}(2012){Kirby}, {Fu}, {Guhathakurta}, \&
  {Deng}}]{Kirby2012}
{Kirby}, E.~N., {Fu}, X., {Guhathakurta}, P., \& {Deng}, L. 2012, \apjl, 752,
  L16

\bibitem[{{Kirby} {et~al.}(2016){Kirby}, {Guhathakurta}, {Zhang}, {Hong},
  {Guo}, {Guo}, {Cohen}, \& {Cunha}}]{Kirby2016}
{Kirby}, E.~N., {Guhathakurta}, P., {Zhang}, A.~J., {et~al.} 2016, \apj, 819,
  135

\bibitem[{{Kumar} {et~al.}(2020){Kumar}, {Reddy}, {Campbell}, {Maben}, {Zhao},
  \& {Ting}}]{Kumar2020}
{Kumar}, Y.~B., {Reddy}, B.~E., {Campbell}, S.~W., {et~al.} 2020, Nature
  Astronomy, 4, 1059

\bibitem[{{Kumar} {et~al.}(2011){Kumar}, {Reddy}, \& {Lambert}}]{Kumar2011}
{Kumar}, Y.~B., {Reddy}, B.~E., \& {Lambert}, D.~L. 2011, \apjl, 730, L12

\bibitem[{{Kumar} {et~al.}(2015){Kumar}, {Reddy}, {Muthumariappan}, \&
  {Zhao}}]{Kumar2015}
{Kumar}, Y.~B., {Reddy}, B.~E., {Muthumariappan}, C., \& {Zhao}, G. 2015, \aap,
  577, A10

\bibitem[{{Li} {et~al.}(2024){Li}, {Shi}, {Li}, {Yan}, \& {Zhang}}]{Li2024}
{Li}, X.-F., {Shi}, J.-R., {Li}, Y., {Yan}, H.-L., \& {Zhang}, J.-H. 2024,
  \mnras, 529, 1423

\bibitem[{{Liu} {et~al.}(2014){Liu}, {Tan}, {Wang}, {Zhao}, {Sato}, {Takeda},
  \& {Li}}]{Liu2014}
{Liu}, Y.~J., {Tan}, K.~F., {Wang}, L., {et~al.} 2014, \apj, 785, 94

\bibitem[{{Magrini} {et~al.}(2021){Magrini}, {Lagarde}, {Charbonnel},
  {Franciosini}, {Randich}, {Smiljanic}, {Casali}, {Viscasillas V{\'a}zquez},
  {Spina}, {Biazzo}, {Pasquini}, {Bragaglia}, {Van der Swaelmen},
  {Tautvai{\v{s}}ien{\.{e}}}, {Inno}, {Sanna}, {Prisinzano}, {Degl'Innocenti},
  {Prada Moroni}, {Roccatagliata}, {Tognelli}, {Monaco}, {de Laverny},
  {Delgado-Mena}, {Baratella}, {D'Orazi}, {Vallenari}, {Gonneau}, {Worley},
  {Jim{\'e}nez-Esteban}, {Jofre}, {Bensby}, {Fran{\c{c}}ois}, {Guiglion},
  {Bayo}, {Jeffries}, {Binks}, {Gilmore}, {Damiani}, {Korn}, {Pancino},
  {Sacco}, {Hourihane}, {Morbidelli}, \& {Zaggia}}]{Magrini2021}
{Magrini}, L., {Lagarde}, N., {Charbonnel}, C., {et~al.} 2021, \aap, 651, A84

\bibitem[{{Mallick} {et~al.}(2022){Mallick}, {Reddy}, \&
  {Muthumariappan}}]{Mallick2022}
{Mallick}, A., {Reddy}, B.~E., \& {Muthumariappan}, C. 2022, \mnras, 511, 3741

\bibitem[{{Martell} {et~al.}(2021){Martell}, {Simpson}, {Balasubramaniam},
  {Buder}, {Sharma}, {Hon}, {Stello}, {Ting}, {Asplund}, {Bland-Hawthorn}, {De
  Silva}, {Freeman}, {Hayden}, {Kos}, {Lewis}, {Lind}, {Zucker}, {Zwitter},
  {Campbell}, {{\v{C}}otar}, {Horner}, {Montet}, \& {Wittenmyer}}]{Martell2021}
{Martell}, S.~L., {Simpson}, J.~D., {Balasubramaniam}, A.~G., {et~al.} 2021,
  \mnras, 505, 5340

\bibitem[{{Meduri} {et~al.}(2024){Meduri}, {Jouve}, \&
  {Ligni{\`e}res}}]{Meduri2024}
{Meduri}, D.~G., {Jouve}, L., \& {Ligni{\`e}res}, F. 2024, \aap, 683, A12

\bibitem[{{Melo} {et~al.}(2005){Melo}, {de Laverny}, {Santos}, {Israelian},
  {Randich}, {Do Nascimento}, \& {de Medeiros}}]{Melo2005}
{Melo}, C.~H.~F., {de Laverny}, P., {Santos}, N.~C., {et~al.} 2005, \aap, 439,
  227

\bibitem[{{Molaro} {et~al.}(2023){Molaro}, {Aguado}, {Caffau}, {Allende
  Prieto}, {Bonifacio}, {Gonz{\'a}lez Hern{\'a}ndez}, {Rebolo}, {Zapatero
  Osorio}, {Cristiani}, {Pepe}, {Santos}, {Alibert}, {Cupani}, {Di
  Marcantonio}, {D'Odorico}, {Lovis}, {Martins}, {Milakovi{\'c}}, {Murphy},
  {Nunes}, {Schmidt}, {Sousa}, {Sozzetti}, \& {Su{\'a}rez
  Mascare{\~n}o}}]{Molaro2023}
{Molaro}, P., {Aguado}, D.~S., {Caffau}, E., {et~al.} 2023, \aap, 679, A72

\bibitem[{{Mori} {et~al.}(2021){Mori}, {Kusakabe}, {Balantekin}, {Kajino}, \&
  {Famiano}}]{Mori2021}
{Mori}, K., {Kusakabe}, M., {Balantekin}, A.~B., {Kajino}, T., \& {Famiano},
  M.~A. 2021, \mnras, 503, 2746

\bibitem[{{Moyano} {et~al.}(2023){Moyano}, {Eggenberger}, {Mosser}, \&
  {Spada}}]{Moyano2023}
{Moyano}, F.~D., {Eggenberger}, P., {Mosser}, B., \& {Spada}, F. 2023, \aap,
  673, A110

\bibitem[{{Palacios} {et~al.}(2006){Palacios}, {Charbonnel}, {Talon}, \&
  {Siess}}]{Palacios2006}
{Palacios}, A., {Charbonnel}, C., {Talon}, S., \& {Siess}, L. 2006, \aap, 453,
  261

\bibitem[{{Plets} {et~al.}(1997){Plets}, {Waelkens}, {Oudmaijer}, \&
  {Waters}}]{Plets1997}
{Plets}, H., {Waelkens}, C., {Oudmaijer}, R.~D., \& {Waters}, L.~B.~F.~M. 1997,
  \aap, 323, 513

\bibitem[{{Rebull} {et~al.}(2015){Rebull}, {Carlberg}, {Gibbs}, {Deeb},
  {Larsen}, {Black}, {Altepeter}, {Bucksbee}, {Cashen}, {Clarke}, {Datta},
  {Hodgson}, \& {Lince}}]{Rebull2015}
{Rebull}, L.~M., {Carlberg}, J.~K., {Gibbs}, J.~C., {et~al.} 2015, \aj, 150,
  123

\bibitem[{{Reddy} \& {Lambert}(2005)}]{Reddy2005}
{Reddy}, B.~E. \& {Lambert}, D.~L. 2005, \aj, 129, 2831

\bibitem[{{Reddy} {et~al.}(2002){Reddy}, {Lambert}, {Hrivnak}, \&
  {Bakker}}]{Reddy2002}
{Reddy}, B.~E., {Lambert}, D.~L., {Hrivnak}, B.~J., \& {Bakker}, E.~J. 2002,
  \aj, 123, 1993

\bibitem[{{Sackmann} \& {Boothroyd}(1999)}]{Sackmann1999}
{Sackmann}, I.~J. \& {Boothroyd}, A.~I. 1999, \apj, 510, 217

\bibitem[{{Sayeed} {et~al.}(2024){Sayeed}, {Ness}, {Montet}, {Cantiello},
  {Casey}, {Buder}, {Bedell}, {Breivik}, {Metzger}, {Martell}, \&
  {McGee-Gold}}]{Sayeed2024}
{Sayeed}, M., {Ness}, M.~K., {Montet}, B.~T., {et~al.} 2024, \apj, 964, 42

\bibitem[{{Schwab}(2020)}]{Schwab2020}
{Schwab}, J. 2020, \apjl, 901, L18

\bibitem[{{Siess} \& {Livio}(1999)}]{Siess1999}
{Siess}, L. \& {Livio}, M. 1999, \mnras, 308, 1133

\bibitem[{{Silva Aguirre} {et~al.}(2014){Silva Aguirre}, {Ruchti}, {Hekker},
  {Cassisi}, {Christensen-Dalsgaard}, {Datta}, {Jendreieck}, {Jessen-Hansen},
  {Mazumdar}, {Mosser}, {Stello}, {Beck}, \& {de Ridder}}]{Silva-Aguirre2014}
{Silva Aguirre}, V., {Ruchti}, G.~R., {Hekker}, S., {et~al.} 2014, \apjl, 784,
  L16

\bibitem[{{Singh} {et~al.}(2021){Singh}, {Reddy}, {Campbell}, {Kumar}, \&
  {Vrard}}]{Singh2021}
{Singh}, R., {Reddy}, B.~E., {Campbell}, S.~W., {Kumar}, Y.~B., \& {Vrard}, M.
  2021, \apjl, 913, L4

\bibitem[{{Smiljanic} {et~al.}(2018){Smiljanic}, {Franciosini}, {Bragaglia},
  {Tautvai{\v{s}}ien{\.{e}}}, {Fu}, {Pancino}, {Adibekyan}, {Sousa}, {Randich},
  {Montalb{\'a}n}, {Pasquini}, {Magrini}, {Drazdauskas}, {Garc{\'\i}a},
  {Mathur}, {Mosser}, {R{\'e}gulo}, {de Assis Peralta}, {Hekker}, {Feuillet},
  {Valentini}, {Morel}, {Martell}, {Gilmore}, {Feltzing}, {Vallenari},
  {Bensby}, {Korn}, {Lanzafame}, {Recio-Blanco}, {Bayo}, {Carraro}, {Costado},
  {Frasca}, {Jofr{\'e}}, {Lardo}, {de Laverny}, {Lind}, {Masseron}, {Monaco},
  {Morbidelli}, {Prisinzano}, {Sbordone}, \& {Zaggia}}]{Smiljanic2018}
{Smiljanic}, R., {Franciosini}, E., {Bragaglia}, A., {et~al.} 2018, \aap, 617,
  A4

\bibitem[{{Sneden} {et~al.}(2022){Sneden}, {Af{\c{s}}ar}, {Bozkurt},
  {Adam{\'o}w}, {Mallick}, {Reddy}, {Janowiecki}, {Mahadevan}, {Bowler},
  {Hawkins}, {Lind}, {Dupree}, {Ninan}, {Nagarajan}, {Topcu}, {Froning},
  {Bender}, {Terrien}, {Ramsey}, \& {Mace}}]{Sneden2022}
{Sneden}, C., {Af{\c{s}}ar}, M., {Bozkurt}, Z., {et~al.} 2022, \apj, 940, 12

\bibitem[{{Soares-Furtado} {et~al.}(2021){Soares-Furtado}, {Cantiello},
  {MacLeod}, \& {Ness}}]{SoarezFurtado2021}
{Soares-Furtado}, M., {Cantiello}, M., {MacLeod}, M., \& {Ness}, M.~K. 2021,
  \aj, 162, 273

\bibitem[{{Stephan} {et~al.}(2020){Stephan}, {Naoz}, {Gaudi}, \&
  {Salas}}]{Stephan2020}
{Stephan}, A.~P., {Naoz}, S., {Gaudi}, B.~S., \& {Salas}, J.~M. 2020, \apj,
  889, 45

\bibitem[{{Strassmeier} {et~al.}(2015){Strassmeier}, {Carroll}, {Weber}, \&
  {Granzer}}]{Strassmeier2015}
{Strassmeier}, K.~G., {Carroll}, T.~A., {Weber}, M., \& {Granzer}, T. 2015,
  \aap, 574, A31

\bibitem[{{Susmitha} {et~al.}(2024){Susmitha}, {Mallick}, \&
  {Reddy}}]{Susmitha2024}
{Susmitha}, A., {Mallick}, A., \& {Reddy}, B.~E. 2024, \apj, 966, 109

\bibitem[{{Takeda} \& {Tajitsu}(2017)}]{Takeda2017}
{Takeda}, Y. \& {Tajitsu}, A. 2017, \pasj, 69, 74

\bibitem[{{Tayar} {et~al.}(2023){Tayar}, {Carlberg}, {Aguilera-G{\'o}mez}, \&
  {Sayeed}}]{Tayar2023}
{Tayar}, J., {Carlberg}, J.~K., {Aguilera-G{\'o}mez}, C., \& {Sayeed}, M. 2023,
  \aj, 166, 60

\bibitem[{{Wallerstein} \& {Sneden}(1982)}]{Wallerstein1982}
{Wallerstein}, G. \& {Sneden}, C. 1982, \apj, 255, 577

\bibitem[{{Yan} \& {Shi}(2022)}]{Yan2022}
{Yan}, H.-l. \& {Shi}, J.-r. 2022, \caa, 46, 1

\bibitem[{{Yan} {et~al.}(2018){Yan}, {Shi}, {Zhou}, {Chen}, {Li}, {Zhang},
  {Bi}, {Wu}, {Li}, {Guo}, {Liu}, {Gao}, {Zhang}, {Zhou}, {Li}, \&
  {Zhao}}]{Yan2018}
{Yan}, H.-L., {Shi}, J.-R., {Zhou}, Y.-T., {et~al.} 2018, Nature Astronomy, 2,
  790

\bibitem[{{Yan} {et~al.}(2021){Yan}, {Zhou}, {Zhang}, {Li}, {Gao}, {Shi},
  {Zhao}, {Aoki}, {Matsuno}, {Li}, {Xu}, {Li}, {Wu}, {Jin}, {Mosser}, {Bi},
  {Fu}, {Pan}, {Suda}, {Liu}, {Zhao}, \& {Liang}}]{Yan2021}
{Yan}, H.-L., {Zhou}, Y.-T., {Zhang}, X., {et~al.} 2021, Nature Astronomy, 5,
  86

\bibitem[{{Zhang} {et~al.}(2020){Zhang}, {Jeffery}, {Li}, \& {Bi}}]{Zhang2020}
{Zhang}, X., {Jeffery}, C.~S., {Li}, Y., \& {Bi}, S. 2020, \apj, 889, 33

\bibitem[{{Zhou} {et~al.}(2022){Zhou}, {Wang}, {Yan}, {Huang}, {Zhang}, {Ting},
  {Zhang}, \& {Shi}}]{Zhou2022}
{Zhou}, Y., {Wang}, C., {Yan}, H., {et~al.} 2022, \apj, 931, 136

\bibitem[{{Zuckerman} {et~al.}(1995){Zuckerman}, {Kim}, \&
  {Liu}}]{Zuckerman1995}
{Zuckerman}, B., {Kim}, S.~S., \& {Liu}, T. 1995, \apjl, 446, L79

\end{thebibliography}

\end{document}